\documentclass{nature} 

\usepackage{graphicx}
\usepackage{caption}
\usepackage{authblk}
\usepackage{hyperref}
\usepackage{float}
\restylefloat{table}

\usepackage{amsmath,amssymb,amsxtra,amstext,amsfonts,dsfont}
\usepackage{color}
\usepackage[normalem]{ulem} 
\usepackage{multirow}
\usepackage{array}
\usepackage{upgreek}
\usepackage{mathrsfs}
\usepackage{mathtools}

\usepackage[squaren]{SIunits}

\newcommand{\bra}[1]{\ensuremath{\langle{#1}|}}
\newcommand{\ket}[1]{\ensuremath{|{#1}\rangle}}
\newcommand{\bracket}[2]{\ensuremath{\langle{#1}|{#2}\rangle}}
\newcommand{\abs}[1]{\ensuremath{\left|{#1}\right|}}
\newcommand{\spl}{\sigma^{+}}
\newcommand{\smn}{\sigma^{-}}
\newcommand{\adg}{a^\dagger}

\newcommand{\HJC}{H_{\text{JC}}}

\newcommand\mr{\mathrm}

\newcommand{\Meff}{M_\text{eff}}
\newcommand{\xeff}{x_\text{eff}}

\title{Schr\"odinger cat states of a 16-microgram mechanical oscillator}

\author[1, 2, $\dagger$]{Marius Bild}
\author[1, 2, $\dagger$, *]{Matteo Fadel}
\author[1, 2, $\dagger$]{Yu Yang}
\author[1, 2]{Uwe von Lüpke}
\author[1, 2]{Phillip Martin}
\author[1, 2]{Alessandro Bruno}
\author[1, 2, **]{Yiwen Chu}

\affil[1]{Department of Physics, ETH Zürich, 8093 Zürich, Switzerland}
\affil[2]{Quantum Center, ETH Zürich, 8093 Zürich, Switzerland}

\affil[*]{fadelm@phys.ethz.ch}
\affil[**]{yiwen.chu@phys.ethz.ch}

\affil[$\dagger$]{these authors contributed equally to this work}
\date{\today}

\begin{document}

\maketitle
\newline
\begin{abstract}
The superposition principle is one of the most fundamental principles of quantum mechanics. According to the Schr\"odinger equation, a physical system can be in any linear combination of its possible states. While the validity of this principle is routinely validated for microscopic systems, it is still unclear why we do not observe macroscopic objects to be in superpositions of states that can be distinguished by some classical property. Here we demonstrate the preparation of a mechanical resonator with an effective mass of 16.2 micrograms in Schr\"odinger cat states of motion, where the constituent atoms are in a superposition of oscillating with two opposite phases. We show control over the size and phase of the superposition and investigate the decoherence dynamics of these states. Apart from shedding light at the boundary between the quantum and the classical world, our results are of interest for quantum technologies, as they pave the way towards continuous-variable quantum information processing and quantum metrology with mechanical resonators. 
\end{abstract}
\maketitle

\clearpage
\newpage
Quantum mechanics is one of the most successful scientific theories ever formulated. However, from the early days of quantum mechanics until now, it has been unclear why quantum phenomena, such as state superpositions, are never observed in the macroscopic world. In his 1935 work \cite{SchroedingerCat35}, Erwin Schr\"{o}dinger imagined a device able to poison a cat as a consequence of a radioactive decay, concluding that the superposition of an atom being ``decayed'' and ``not decayed'' could be mapped onto a superposition of the cat being simultaneously ``dead'' and ``alive''. There are two aspects of this hypothetical scenario that make it seem absurd and counter intuitive: First, a cat is a macroscopic, everyday object, and second, "dead" and "alive" are states with properties that are clearly distinguishable within our classical experience.

Many explanations have been proposed as to why we may never encounter a cat in such an unfortunate situation. Macroscopic objects may simply be too complex and subject to too many sources of decoherence to sustain a superposition of classically distinct states. Other theories introduce additional effects beyond standard quantum mechanics, such as wavefunction collapse due to intrinsic stochastic noise or gravitational decoherence\cite{BassiRMP13}. In keeping with the spirit of Schr\"{o}dinger's cat, these effects are typically expected to scale with the mass of the system and the distinctness of the states that are superposed. Therefore, observing state superpositions in massive objects is of key importance for exploring the validity range of quantum mechanics as we know it. Beyond its fundamental interest, preparing and detecting Schr\"{o}dinger's cat states is essential for applications in quantum technologies. Main examples include Heisenberg-limited parameter estimation protocols\cite{PezzeBook, Munro2002} and error-protected quantum information processing\cite{CochranePRA99,MirrahimiNJP14}. 

There have been many experimental demonstrations of Schr\"{o}dinger cat states (which we will call ``cat states" from here on). These include superpositions of internal and motional degrees of freedom in trapped ions\cite{MonroeCatSCI96, Lo2015}, phase-space superpositions of electromagnetic waves in both the optical\cite{ourjoumtsev2007, huang2015} and microwave domains\cite{auffeves2003, DelegliseNat08, Vlastakis2013}, Greenberger–Horne–Zeilinger states\cite{leibfried2005, gao2010}, current superpositions in SQUIDs\cite{FriedmanNat00}, and spatial superpositions of large molecules\cite{gerlich2011}. In this work, we experimentally demonstrate the preparation of cat states in the motional degree of freedom of a solid state mechanical resonator.
Given the variety of definitions found in previous works, here we define a cat state of a harmonic oscillator as a coherent superposition of two or more states with well-separated phase space distributions.

\begin{figure}
\centering
\includegraphics[width=8cm]{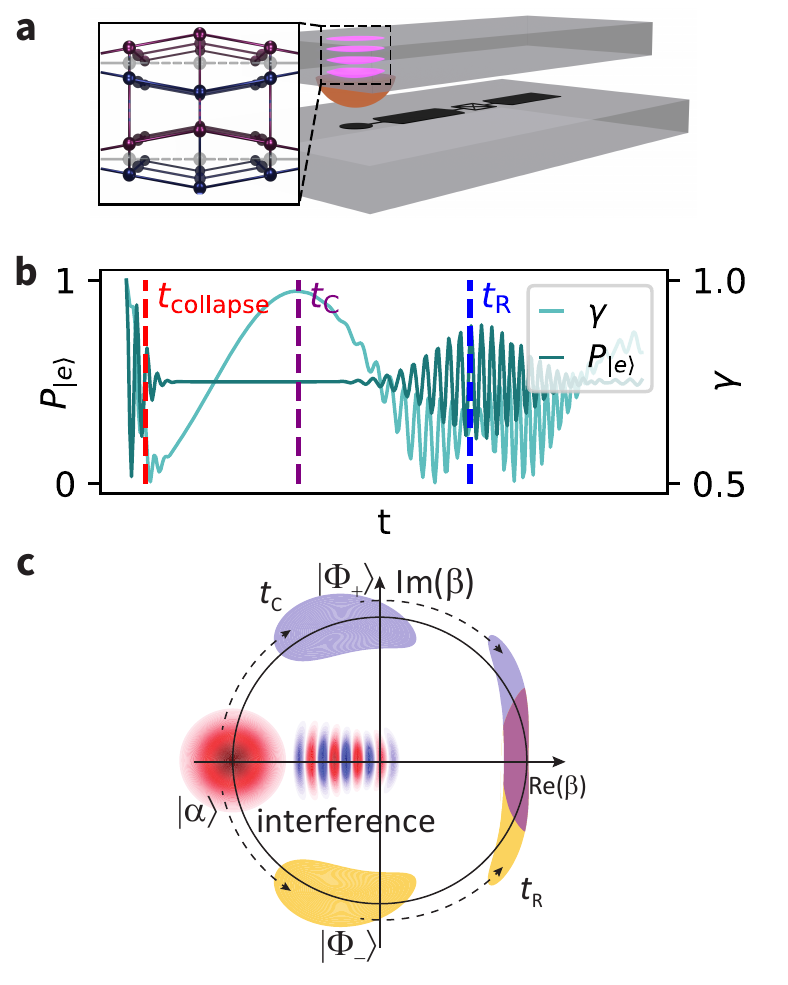}
\caption{\textbf{Illustration of the $\hbar$BAR device and system evolution.}
\textbf{(a)} Schematics of the $\hbar$BAR device. The HBAR chip (top) has a layer of piezoelectric aluminum nitride (orange) and supports standing acoustic waves (pink). The transmon qubit on the lower chip has a circular antenna to couple with the HBAR. The inset shows the superposition of two opposite-phase oscillations of atoms in the crystal lattice.
\textbf{(b)} Simulated evolution of the qubit $\ket{e}$ state population $P_{\ket{e}}$ and purity $\gamma$ under the JC interaction when the qubit is initialized in $\ket{-Z}$ and the phonon in a coherent state. 
\textbf{(c)} Illustration of the evolution of an initial phonon coherent state (red circle on the left) in phase space. The blue (yellow) crescent shapes indicate the state $\ket{\Phi_+}$ ($\ket{\Phi_-}$), which is the phonon state when the qubit is initialized in $\ket{+X}$ ($\ket{-X}$). Interference fringes appear around time $t_C$ when the qubit is prepared in a superposition of $\ket{+X}$ and $\ket{-X}$. Around the revival time $t_R$, the two phonon states again overlap (purple).
}
\label{fig1}
\end{figure}

Our mechanical resonator is a high-overtone bulk acoustic-wave resonator (HBAR), which we couple to a superconducting transmon qubit. The latter allows us to create, control, and read out phonon states in the HBAR. Qubit and HBAR are fabricated on separate sapphire chips, which are subsequently flip chip bonded into the final device  (Fig.~\ref{fig1}a). 
The acoustic free spectral range is approximately $12\,$MHz, and frequency tuning the qubit allows us to address several longitudinal phononic modes.
The acoustic lattice oscillations are localized within a Gaussian mode with waist $w_0 = 27\,\mu$m and length $L = 435\,\mu$m, giving a mode volume of $\pi w_0^2 L\approx 0.001$\,mm$^3$ (see supplementary materials\cite{SM}, section B for details). More details about this circuit quantum acoustodynamics (cQAD) system\cite{Chu2018} and the device\cite{vonLupke22} can be found in previous works.

In the classical picture, one can imagine a coherent state $\ket{\alpha}$ in the phonon mode as a coherent displacement of the atomic lattice with an amplitude proportional to $\alpha$. In the quantum picture, an example of a cat state is a quantum superposition of two coherent states with opposite displacement amplitudes, leading to the physical interpretation of such a state as the superposition of two oscillations of the atomic lattice with the same frequency $\omega_p$ and relative phase $\pi$.
Considering a snapshot in time where both oscillations are at their displacement maximum, Schr\"odinger's cat being in a superposition of dead and alive is analogous to a superposition of atoms in the HBAR being in two distinct positions in space, as illustrated  in the inset of Fig.\ref{fig1}a. Note that here, we define the positions as distinct when their separation is larger than the fluctuations due to quantum, thermal, or other sources of noise. 

To realize a cat state in our system, we use the Jaynes-Cummings (JC) interaction with the qubit and phonon on resonance\cite{BuzekPRA92, auffeves2003}. The interaction Hamiltonian is
\begin{equation}
    H/\hbar = g_0 ( \spl a + \smn\adg ) \;, \label{eq:HJC}
\end{equation}
where $g_0$ is the coupling strength between qubit and phonon mode, $\sigma^+$ is the raising operator for the qubit, and $a^\dagger$ the raising operator for the phonon mode. This results in Rabi oscillations between the states $\ket{e, n-1}$ and $\ket{g, n}$ at a rate $g_0\sqrt{n}$, where $\ket{g}$ ($\ket{e}$) is the qubit ground (excited) state and $\ket{n}$ is the Fock state of $n$ phonons. As a consequence of this $\sqrt{n}$ scaling, if the phonon mode is prepared in a coherent state with large enough amplitude and the qubit is prepared in $\ket{g}$ or $\ket{e}$, their coherent interaction rapidly dephases. Hence, the oscillations of the qubit population "collapses" (see Fig. \ref{fig1}b) with a decaying amplitude proportional to\cite{Cummings65, Rempe87} $\text{exp}(-(t/t_{\text{collapse}})^2)$, where $t_{\text{collapse}}=\sqrt{2}/g_0$ is the collapse time in the limit of $\alpha \gg 1$. At this time, the qubit and phonon states are entangled. This can be seen in Fig.\ref{fig1}b as a minimum in the qubit state purity $\gamma(t)=\text{Tr}\left(\rho_\text{q}(t)^2\right)$ around $t_{\text{collapse}}$, where $\rho_\text{q}(t)$ is the reduced density matrix of the qubit. Strikingly, due to the quantized phonon energy and the consequent discrete oscillation frequency spectrum, the oscillations revive in finite time\cite{BuzekPRA92}. For $\alpha \gg 1$, this revival occurs at $t_R = 2 \pi \alpha/g_0$\cite{EberlyPRL80}. 
Between the collapse and revival, at time $t_R/2$, the qubit and phonon disentangle from each other. The state being separable at $t_R/2$ coincides with the occurrence of a superposition of two distinct states in phase space, realizing a cat state in the phonon mode\cite{SM, BuzekPRA92,GeaPRL90, auffeves2003}.            

A more intuitive explanation for the origin of the cat state comes from the time evolution of the reduced phonon state in phase space. As illustrated in Fig.~\ref{fig1}c and shown in the supplementary materials\cite{SM}, section A, if the qubit is initialized in the state $\ket{\pm X} \equiv (\ket{e}\pm\ket{g})/\sqrt{2}$ and in the limit of large $\alpha$, the evolution leads to a rotation in phase space with an angular velocity $\mp\abs{g_0/2\alpha}$ and a distortion of the coherent states. We call the resulting states $\ket{\Phi_\pm (t)}$, whose full expressions are given in the supplementary materials\cite{SM} (Eq.~S15). Initializing the qubit in the state $\ket{\pm Z} \equiv (\ket{+X}\pm\ket{-X})/\sqrt{2}$, the phonon state will evolve into $\ket{\Phi_+(t)}\pm\ket{\Phi_-(t)}$ as shown in Fig.~\ref{fig1}c. At time $t_R/2$, the two state components $\ket{\Phi_\pm}$ have covered a rotation angle of $\mp\pi/2$ around a circle of radius $\alpha$, maximizing their separation in phase space and forming a cat state\cite{BuzekPRA92}. Finally, at the revival time $t_R$, the two phonon state components $\ket{\Phi_\pm}$ have both rotated by a phase of $\pi$ and approximately recombine in phase space.

\begin{figure}
\centering
\includegraphics[width=8cm]{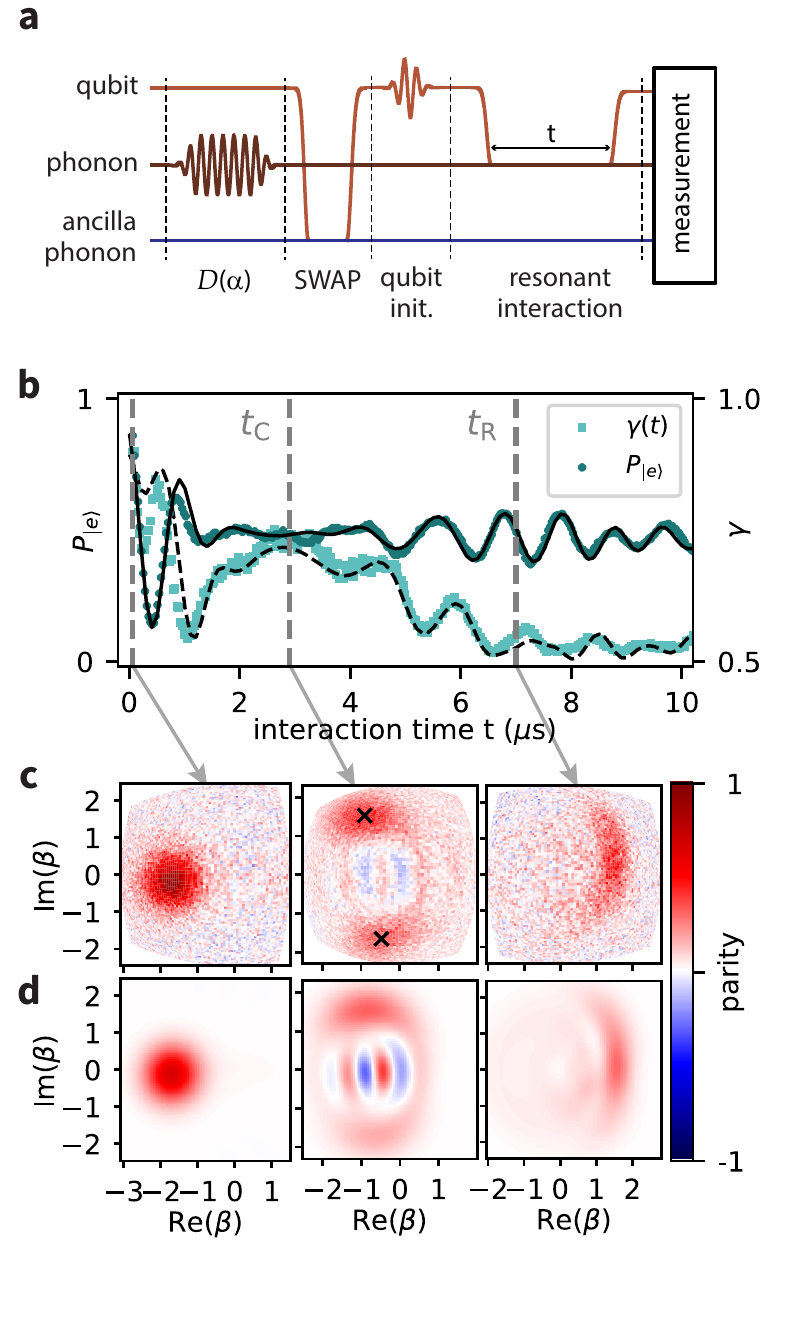}
\caption{\textbf{Collapse and revival dynamics.}
\textbf{(a)} Experimental sequence for observing collapse and revivals dynamics and for preparing cat states (details in the main text).
\textbf{(b)} Measured qubit population and state purity. The solid and dashed black lines are the simulation results of the qubit population and purity, respectively. Three time points of particular interest are highlighted (dashed lines): initial state time, cat state time ($t_C$), and revival state time ($t_R$). \textbf{(c)} Measured Wigner function of the phonon state at the three time points. Axes are the real and imaginary parts of the complex displacement amplitude $\beta$ used during Wigner tomography\cite{vonLupke22}. The black crosses indicate the positions of the two coherent states composing the fitted CSS state Eq.~\eqref{eq: ideal_cat}. \textbf{(d)} Corresponding simulated Wigner functions.
\label{fig2} }
\end{figure}

In the following, we experimentally confirm both the predicted collapse and revival of Rabi oscillations and the creation of mechanical cat states in the phonon mode.
The basic sequence used in the experimental demonstration of the JC dynamics described above can be seen in Fig.~\ref{fig2}a.    
We displace the phonon mode with a resonant drive of amplitude $A$ to a coherent state with amplitude $\alpha$. To mitigate any effect of the drive on the qubit state, we then cool the qubit with an ancillary phonon mode\cite{Chu2018, vonLupke22}. The qubit is subsequently prepared in its initial state by applying a drive pulse with variable phase and amplitude. In order to induce the resonant interaction, we tune the qubit to the phonon mode frequency for a variable interaction time $t$.
Depending on which of the subsystems we want to characterize, we choose a measurement sequence that implements the appropriate measurement operator. First, we simply measure the qubit excited state population. The resulting data is shown in Fig.~\ref{fig2}b for $A=0.35$. Here the value for $A$ is a scaling factor for the amplitude of a microwave drive, which we calibrate to find a corresponding initial coherent state size of $\alpha=1.75$ (see supplementary materials\cite{SM}, section D). As expected, we observe oscillations that collapse after a time $t_{\text{collapse}}\approx\unit{0.9}{\mu s}$ and revive at $t_R\approx\unit{6.7}{\mu s}$. This revival indicates the coherent exchange of energy quanta between the qubit and phonon mode during the resonant interaction. By performing full qubit tomography after the resonant interaction, we can also reconstruct the reduced density matrix of the qubit subsystem $\rho_{\text{q}}$ and calculate the purity of the qubit state $\gamma(t)$. We confirm a local minimum of $\gamma(t)$ around the predicted collapse time $t_{\text{collapse}}$, followed by a local maximum around $t_R/2$ (Fig.~\ref{fig2}b). 

We now focus on the time evolution of the phonon subsystem by performing full Wigner tomography of the phonon state after the resonant interaction times $t=0,\, 2.9$ and $7.0\,\mu$s. To this end, we use the parity measurement technique established in a previous work\cite{vonLupke22}.
To compensate for the effect of qubit dephasing during the parity measurement, we normalize all measured parity values to that of the Fock $\ket{0}$ phonon state (see supplementary materials\cite{SM}, section C).
The measured Wigner functions are shown in Fig.~\ref{fig2}c, where axes in phase space are normalized by measuring the distribution of populations in the phonon Fock states for coherent states created with different drive amplitudes\cite{Chu2018} (see supplementary materials\cite{SM}, section D).

From the measured data, we confirm the evolution of the initial coherent state (Fig.~\ref{fig2}c left) into a cat state at $t_C=2.9\,{\mu}$s (Fig.~\ref{fig2}c center), showing two state components clearly distinct in phase space and interference fringes located between them. We choose this value of $t_C$ because it corresponds to the measured maximum in the qubit state purity. It deviates somewhat from the value predicted using the large $\alpha$ limit, which is $t_R/2\approx 3.3\,{\mu}$s. For the evolution time $t=7.0\,\mu$s, the predicted refocusing into a crescent shaped overlap between the counter-rotating state components can be observed (Fig.~\ref{fig2}c right).

In order to benchmark the cat state and obtain an estimate of its size, we implement a maximum-likelihood reconstruction\cite{Chou2018} of the phonon state $\rho_{\text{p}}$ from the measured state with $A=0.35$ and $t_C=2.9\,{\mu s}$. We then fit the reconstructed state to an analytical expression of the expected phonon state $\rho(t_C)$ in the absence of decoherence and after tracing out the qubit (see supplementary materials\cite{SM}, section A). Fixing the interaction time to $t_C$ from the experiment, the fit maximizes the fidelity between $\rho_{\text{p}}$ and $\rho(t_C)$  by varying the initial coherent state size $\alpha_{\text{fit}}$ of $\rho(t_C)$. The result yields a fidelity of $\mathcal{F}\approx76\%$ to an analytical state with initial coherent state size $\alpha_{\text{fit}}=1.62$, which is smaller than the initial displacement $\alpha=1.75$ because the expression for $\rho(t_C)$ does not include phonon losses. We attribute the infidelity to a combination of decoherence and measurement imperfections that lead to additional artifacts in the Wigner function\cite{vonLupke22}. To further confirm that the phonon state behaves as expected, Fig.~\ref{fig2}d shows the results of a master equation simulation of the full experimental protocol with independently measured system parameters, showing good agreement with the measurements in Fig.~\ref{fig2}c.

The state we obtained resembles the two-component coherent state superpositions (CSS)
\begin{equation}
    \ket{C} = \mathcal{N} \left( \ket{\alpha_1} + e^{i\vartheta} \ket{\alpha_2} \right), \label{eq: ideal_cat}
\end{equation} 
a type of cat state that is often invoked in quantum information\cite{CochranePRA99, MirrahimiNJP14} and parameter estimation protocols\cite{PezzeBook, Munro2002}. Here $\ket{\alpha_{1, 2}}$ are two coherent states and $\mathcal{N}$ is the appropriate normalization constant. We can fit our reconstructed state to Eq.~\eqref{eq: ideal_cat} by optimizing $\alpha_1$, $\alpha_2$, and $\vartheta$ for maximum fidelity $\mathcal{F}\left(\rho_\text{p}, \ket{C}\bra{C}\right)$. Since our state is not centered around the origin in phase space, we use half the phase space distance between the coherent state components $D=\vert\alpha_1-\alpha_2\vert/2$ as a measure of the cat state size. This choice is motivated by considering a coherent state superposition centered around the origin in phase space, such that $\alpha_1=-\alpha_2$. Then $D=\vert\alpha_{1,2}\vert=\sqrt{\bar{n}}$, where $\bar{n}$ is the average phonon population of the state created. For the state in Fig.~\ref{fig2}c, we obtain a state size $D=1.61$, corresponding to $\bar{n} = D^2 = 2.60$, with a fidelity of $\mathcal{F}\approx 66\%$. The smaller state size $D$ compared to the initial coherent displacement is a combination of decoherence and the choice of interaction time $t_C<t_R/2$, resulting in the two counter-rotating state components not reaching their maximum separation in phase space. The fidelity is lower compared to the fitted analytical state $\rho(t_C)$, since $\rho(t_C)$ itself has a finite infidelity to the CSS state $\ket{C}$.

We can now translate the parameters of the measured cat state into physical properties of the phonon mode, such as the spatial separation between atoms. A state size of $D=1.61$ corresponds to a maximal delocalization of $7.0\cdot x_{\text{ZPF}}$, where $x_{\text{ZPF}}$ is the zero point motion of an equivalent 1D quantum harmonic oscillator. Since we are not considering a center-of-mass mode, there is some freedom in choosing $x_{\text{ZPF}}$, which is then associated with an effective oscillating mass of the mode. If we choose the root-mean-square (rms) value of the atomic displacements, we find an effective mass of $M^{(\text{rms})}_\text{eff}= 16.2\,\mu$g, corresponding to $\sim10^{17}$ atoms, delocalized over a distance of $2.1\cdot 10^{-18}\,$m (see supplementary materials\cite{SM}, section B).

\begin{figure}
\centering
\includegraphics[width=8cm]{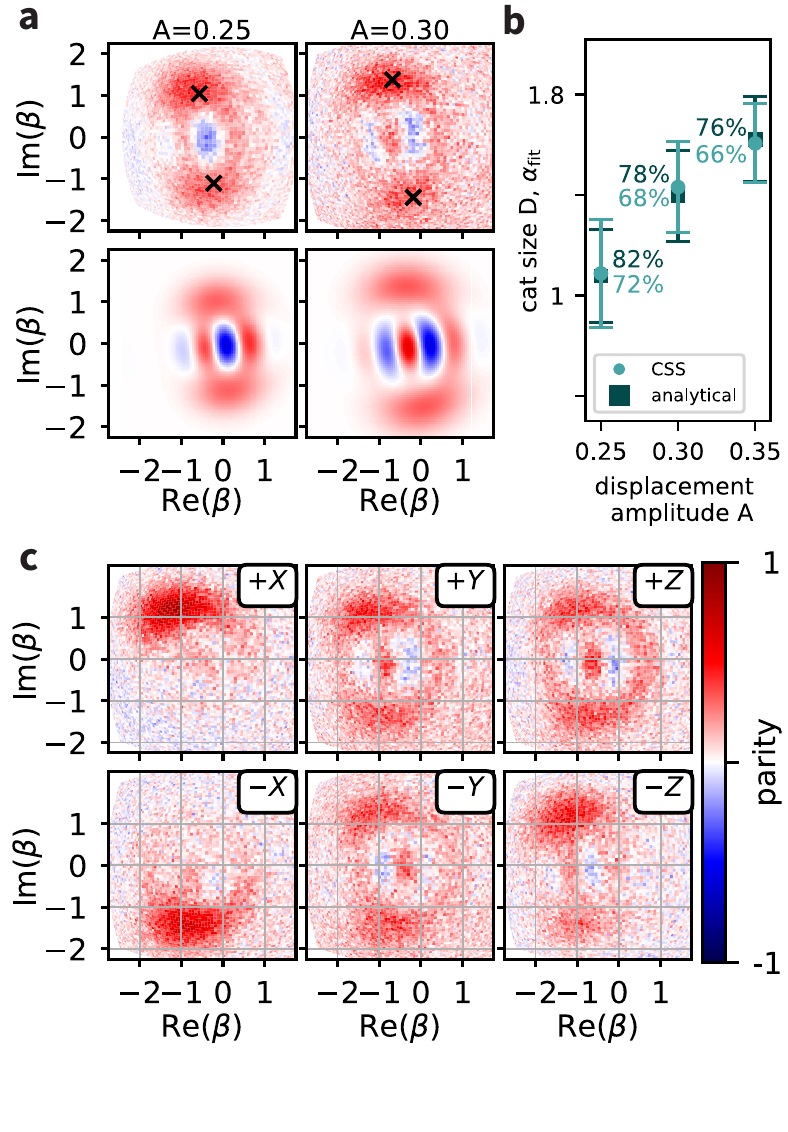}
\caption{\textbf{Cat state amplitude and phase control.}
\textbf{(a)} Cat states prepared with different displacement pulse amplitudes $A$. Top row: measured Wigner functions, Bottom row: Analytical state $\rho(t_C)$ which best fits the data. The fitted CSS states, with coherent state positions indicated by black crosses, have $D$ = 1.09 (1.43) for $A$ = 0.25 (0.30).
\textbf{(b)} Cat state sizes as a function of the displacement amplitude $A$, obtained from fitting the data in \textbf{a} to analytical states $\rho(t_C)$ and CSS states $\ket{C}$. Numbers are the fidelity with respect to the fitted state. Error bars show cat sizes resulting in 1$\%$ deviation in the fidelity (see supplementary materials \cite{SM}, section F)
\textbf{(c)} Cat states resulting from the initial qubit states indicated by the respective labels.
\label{fig3} }
\end{figure}

In applications such as bosonic encodings of a qubit state, full control over the phase and amplitude of the created cat state is required\cite{CochranePRA99, MirrahimiNJP14, Grimm2020}. In the following, we demonstrate this level of control in our experiment.
By varying the amplitude $A$ of the phonon displacement drive, we can control the amplitude of the initial coherent state and the size of the resulting cat state. For displacement amplitudes $A=0.25$ and $0.30$, we create cat states with $D=1.09$ and $1.43$, respectively (Fig.~\ref{fig3}a).
The fidelities of reconstructions of the measured states to both a CSS state and $\rho(t_C)$ are given in Fig.~\ref{fig3}b. The best fit $\rho(t_C)$ are plotted in the lower row of Fig.~\ref{fig3}a, showing good qualitative agreement with the data. As before, the finite fidelities and lower contrast fringes of the measured states compared to the best fit $\rho(t_C)$ arise mainly from decoherence of the state during measurement.

In the two-component cat state encoding of a qubit, the six cardinal points of the Bloch sphere are given by two coherent states $\ket{\alpha_1}$ and $\ket{\alpha_2}$, along with their four superpositions with $\pi/2$ difference in the phase $\vartheta$ (see Eq.\eqref{eq: ideal_cat}). We can prepare similar states by initializing the transmon qubit state in all six cardinal points $\ket{\pm X}, \ket{\pm Y}, \ket{\pm Z}$ of its Bloch sphere before performing the cat generation protocol. The preparation of $\ket{\pm X}$ and $\ket{\pm Y}$ is calibrated using collapse and revival measurements as a function of the qubit drive phase (see supplementary materials\cite{SM}, section A). The results for $A=0.35$ and $t_C=2.10\,\mu$s are shown in Fig. \ref{fig3}c. We observe a distorted coherent state located on the upper (lower) half of phase space for the qubit initially in $\ket{\pm X}$. This separation in phase space is expected from the opposite rotation directions between the phonon states when the qubit is prepared in $\ket{\pm X}$ (see Fig.\ref{fig1}c). The $\ket{\pm Y}, \ket{\pm Z}$ states then give rise to four cat states that differ in phase by $\pi/2$, as can be observed in the phases of the interference fringes in Fig. \ref{fig3}c. Note that the initial energy of the qubit, and thus of the total system, is not the same for all six scenarios, resulting in slightly different sizes for the cat states. In the limit of large cat size, this difference becomes negligible, and the phonon subspace maps onto that of the cat state encoding.

\begin{figure}
\centering
\includegraphics[width=16cm]{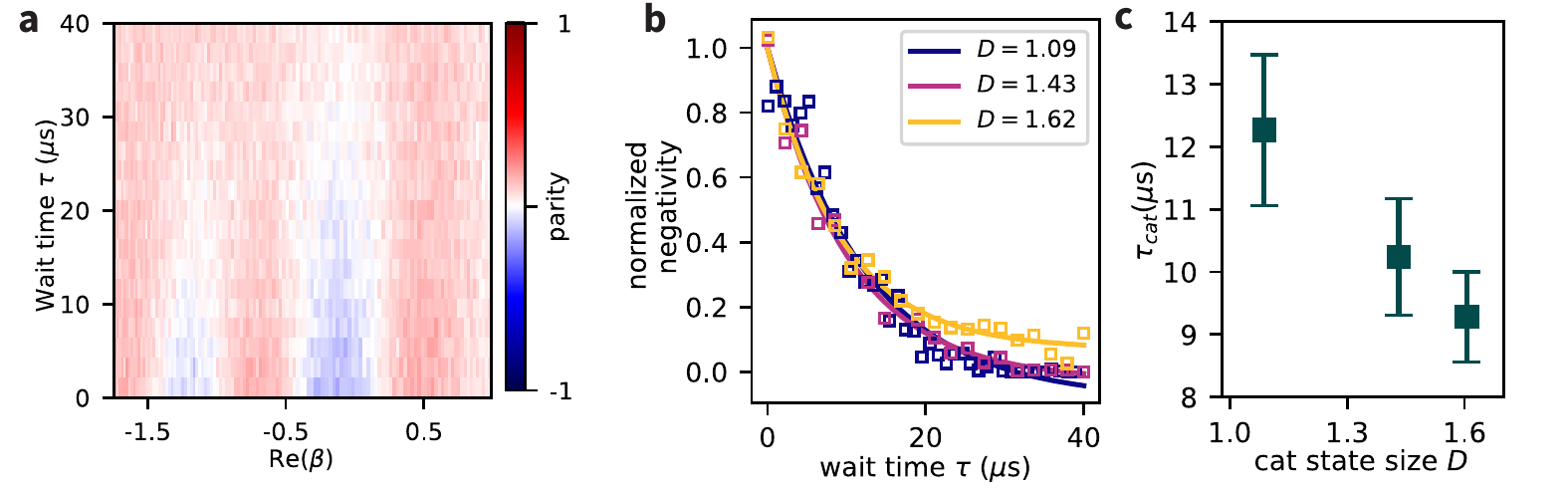}
\caption{\textbf{Decoherence of cat states.}
\textbf{(a)} Measured 1D cuts through the interference fringes of the $D=1.43$ cat state for a range of wait times between state creation and measurement. \textbf{(b)}: Extracted negativities (squares) from each cut vs. wait times for three cat state sizes, together with fitted exponential decays (solid lines). Both data and fitted curves are normalized to the fitted value at $\tau = 0$. \textbf{(c)} Characteristic decay times $\tau_\text{cat}$ extracted from the fits in \textbf{(b)} for all three cat state sizes. Errorbars are uncertainties extracted from the fit.
\label{fig4} }
\end{figure}

Superposition states are non-classical states that are notoriously prone to decoherence. We now investigate the quantum to classical transition of different sized cat states by letting them evolve freely for a varying wait time $\tau$ before performing Wigner tomography. In particular, we focus on a slice through the Wigner function's interference fringes at Im($\beta$) = 0, which highlights the non-classical features of the superposition. Fig.~\ref{fig4}a shows the time evolution of this slice for the $D = 1.43$ cat state. We observe that the negative features disappear on a time scale much faster than $T_1^{\text{ph}} \approx 84\,\mu s$, the energy relaxation time of the phonon mode.

As a measure for the non-classicality of the state, we extract the time dependent negativity\cite{Kenfack04}, defined as $\delta(t) \equiv \int \left(\vert W(\beta, t)\vert - W(\beta, t)\right) d\beta$. Here, $W(\beta, t)$ is the measured Wigner function of the cat state at time $t$, and the integration is over the 1D slice in phase space parametrized by the complex displacement amplitude $\beta$. Fig.~\ref{fig4}b shows the resulting $\delta(t)$ for the three different cat state sizes of Fig.~\ref{fig3}b. We fit each dataset to an exponential decay plus a constant offset. The offset in the measured Wigner values arises from the fact that our Wigner tomography is not performed in the ideal dispersive limit\cite{vonLupke22}. The extracted decay timescales $\tau_\text{cat}$ are plotted in Fig.~\ref{fig4}c. We show in section G of the supplementary materials\cite{SM} that, in the limit of large $\abs{\alpha}$, $\delta(\tau)$ decays exponentially with a time constant $\tau_\text{cat}=T_1^{\text{ph}}/(2\abs{\alpha}^2)$. However, for small $\abs{\alpha}$, $\tau_\text{cat}$ deviates from this expression and is in fact dependent on properties of the exact state, such as the phase of the superposition. The data in Fig.~\ref{fig4}c shows the expected qualitative behavior of faster decaying negativity for larger sized cat states, and we present a more detailed quantitative analysis in the supplementary materials\cite{SM}.

Our results show the generation of cat states in a microgram-mass solid-state mechanical mode using the tools of cQAD. We observed the collapse and revival dynamics of the qubit population under the resonant JC interaction and used it to create cat states of different sizes and superposition phases. 
As a proof-of-principle demonstration of how such macroscopic superpositions can be used to investigate mechanisms of quantum decoherence, we measured the decay of negativity in the Wigner function of different sized cat states. Future tests of wavefunction collapse models using such systems\cite{schrinski_2022} would benefit from larger sized cats and longer phonon lifetimes. We note that the HBAR mode is a standing wave with longitudinal mode number $\sim 500$, and a half-wavelength section  approximates a center-of-mass mode. The mass of this section is on the order of 30 nanograms, by far the most massive object that has been placed into a cat state. 

The maximum size of the cat state we can prepare is currently limited by our device parameters, including both the qubit and phonon decoherence rates. The latter is especially important given that, in general, the decoherence rate of the cat state is proportional to the square of the cat state size $D$. Furthermore, additional improvements to the properties of qubit and phonon resonator would enable alternative cat state generation protocols that can in principle lead to states with a higher fidelity to, for example, a CSS state\cite{DelegliseNat08, Vlastakis2013}. We point out, however, that while CSS states represent a useful benchmark because they have been extensively studied for applications such as quantum information and quantum metrology, many of their salient features are present already in the states we have demonstrated. These include the phase-space separation of state components, important for error protection of encoded qubits\cite{Ofek2016, Grimm2020, Fluhmann2019}, and the presence of interference fringes with high Fisher information, useful for quantum-enhanced sensing~\cite{Munro2002,JooPRL11,Facon16}.

\section*{Acknowledgements}

We thank Oriol Romero-Isart, Alexander Grimm, and Ines C. Rodrigues for useful discussions, and Arianne Brooks for help with figure making. Fabrication of devices was performed at the FIRST cleanroom of ETH Zürich and the BRNC cleanroom of IBM Zürich. M.B. was supported by the QuantERA II Program that has received funding from the European Union’s Horizon 2020 research and innovation program under Grant Agreement No 101017733, and with the Swiss National Science Foundation. M.F. was supported by The Branco Weiss Fellowship -- Society in Science, administered by the ETH Z\"{u}rich. 
\section*{Author contributions statement}

U.v.L fabricated the device. M.B., M.F., Y.Y., and U.v.L performed the experiments and analyzed the data. All authors performed theoretical calculations and simulations. Y.C. conceived of the project and supervised the work. M.B., M.F., Y.Y., U.v.L, and Y.C. wrote the manuscript.

\section*{Additional information}

\textbf{Competing interests} The authors declare no competing interests. \\
\textbf{Data and code availability} Raw data, analysis code and QuTiP simulations are available from the corresponding author on reasonable request. Requests for materials should be addressed to M.F or Y.C.\\

\clearpage
\newpage

\renewcommand{\thesection}{}  
\renewcommand{\thesubsection}{\Alph{subsection}}  

\renewcommand{\thetable}{S\arabic{table}}  
\renewcommand{\thepage}{S\arabic{page}}  
\renewcommand{\thefigure}{S\arabic{figure}}
\renewcommand{\theequation}{S\arabic{equation}}
\setcounter{page}{1}
\setcounter{figure}{0}
\setcounter{table}{0}
\setcounter{section}{0}
\setcounter{equation}{0}


\begin{center}
    \section*{Supplementary information for ``Schr\"odinger cat states of a 16-microgram mechanical oscillator"}
\end{center}
    Marius Bild$^{1,2,\dagger}$, Matteo Fadel$^{1,2,\dagger,*}$, Yu Yang$^{1,2,\dagger}$, Uwe von Lüpke$^{1,2}$,
    Phillip Martin$^{1,2}$,
    Alessandro Bruno$^{1,2}$,
    Yiwen Chu$^{1,2,**}$\\
    $^1$ \textit{Department of Physics, ETH Zürich, 8093 Zürich, Switzerland} \\
    $^2$ \textit{Quantum Center, ETH Zürich, 8093 Zürich, Switzerland} \\
    $^\dagger$ these authors contributed equally to this work\\
    $^*$ fadelm@phys.ethz.ch\\
    $^{**}$ yiwen.chu@phys.ethz.ch
\tableofcontents

\clearpage
\newpage
\subsection{State evolution}\label{si:state_evolution}
\hfill\\
For a resonant interaction between the qubit and the phonon, we can consider the time evolution given by the Jaynes–Cummings Hamiltonian
\begin{equation}
    \HJC = g_0 ( \spl a + \smn\adg ) \,.
\end{equation}
Consider the initial product state
\begin{equation}
    \ket{\Psi_0} = \left(c_g \ket{g} + c_e \ket{e} \right) \otimes \ket{\alpha} \;,
\end{equation}
where
\begin{equation}
    \ket{\alpha} = e^{-\vert\alpha\vert^2/2} \sum_{n=0}^\infty \dfrac{\alpha^n}{\sqrt{n!}} \ket{n} \equiv \sum_n c_n \ket{n}
\end{equation}
is a coherent state for the field. In the following, we will consider $\alpha\in\mathbb{R}$ for simplicity.

Defining the time-evolution operator $U_{JC}(t) = e^{-i \HJC t}$ we get
\begin{align}
    U_{JC}(t) \ket{g,n} &= \cos(g_0 \sqrt{n} t) \ket{g,n} - i \sin(g_0 \sqrt{n} t) \ket{e,n-1} \\
    U_{JC}(t) \ket{e,n} &= \cos(g_0 \sqrt{n+1} t) \ket{e,n} - i \sin(g_0 \sqrt{n+1} t) \ket{g,n+1},
\end{align}
which gives
{\small
\begin{align} 
    \ket{\Psi(t)} &= U_{JC}(t) \ket{\Psi_0} \\
    &= \sum_{n=0}^\infty c_n \left[ c_g (\cos(g_0\sqrt{n} t)\ket{g,n} - i \sin(g_0\sqrt{n}t)\ket{e,n-1} )\right.\\
    &\hspace{18pt}+\left. c_e (\cos(g_0\sqrt{n+1} t)\ket{e,n} - i \sin(g_0\sqrt{n+1}t)\ket{g,n+1} ) \right] \notag\\
    &= \sum_{n=0}^\infty \left[ (c_n c_g \cos(g_0\sqrt{n} t) - i c_{n-1} c_e \sin(g_0 \sqrt{n} t) ) \ket{g} \right.\label{eq:sup_exactState}\\
    &\hspace{18pt}+\left.(c_n c_e \cos(g_0\sqrt{n+1} t) - i c_{n+1} c_g \sin(g_0\sqrt{n+1} t) ) \ket{e}
    \right] \ket{n}\notag \;.
\end{align}
}
\clearpage
Tracing out the qubit state, we are left with the phonon state density matrix 
\begin{equation}
    \rho(t) = \text{Tr}_{\text{q}}\left(\ket{\Psi(t)}\bra{\Psi(t)}\right) = \sum_{i\in \{e,g\}}\bracket{i}{\Psi(t)}\bracket{\Psi(t)}{i}\,.\label{eq:sup_exactDM}
\end{equation}
This is the ``analytical state'' that we then use to fit to the reconstructed measured states.

Inserting $c_{n+1}=c_n \alpha/\sqrt{n+1}$, the probability to measure the qubit in the excited state is
{\small
\begin{align}
    P_e (t) = \sum_{n=0}^\infty & \abs{c_n}^2 \left( \abs{c_e}^2 \cos(g_0\sqrt{n+1} t)^2 + \abs{c_g}^2 \dfrac{\abs{\alpha}^2}{n+1} \sin(g_0\sqrt{n+1} t)^2 \right) \notag\\
    & + 2 \cos(g_0\sqrt{n+1} t)\sin(g_0\sqrt{n+1} t) \mathfrak{Im}[c_{n+1}c_n^\ast c_g c_e^\ast] \;. \label{eq:sup_peExact}
\end{align}
}
Note that, for $\alpha=0$, only the term $n=0$ matters, and one recovers the expression for the vacuum Rabi oscillations $P_e (t) = \abs{c_e}^2 \cos(g_0 t)^2$.
In the limit of $\alpha\gg 1$ and $t \ll \alpha^2/g_0$, the sum in Eq.~\eqref{eq:sup_peExact} can be well approximated by a closed form. To see this, note that for $\alpha\gg 1$ the distribution of $c_n$ is sharply peaked around $n \approx \alpha^2$. This allows us to expand $g_0t\sqrt{n+1}\approx g_0t\sqrt{n}$ as 
\begin{align}\label{eq:sup_sqrtNapprox}
    g_0t\sqrt{n} &= g_0t\sqrt{\alpha^2+(n-\alpha^2)} \notag\\
    &\approx g_0t\left(\alpha + \dfrac{1}{2}\dfrac{(n-\alpha^2)}{\alpha} - \dfrac{1}{8}\dfrac{(n-\alpha^2)^2}{\alpha^3} \right)\;,
\end{align}
where terms of even higher order in $n$ can be neglected for $g_0t \ll 8 \alpha^2$, since the standard deviation of $(n-\alpha^2)$ is equal to $\alpha$ for a Poisson distribution.

If we consider only the term $\alpha$ in the expansion Eq.~\eqref{eq:sup_sqrtNapprox}, which is valid for short times $g_0t \ll 2$ or for the interaction with a classical field, we obtain the expression for undamped Rabi oscillations $P_e (t) = \abs{c_e}^2 \cos(g_0 \alpha t)^2 + \abs{c_g}^2 \sin(g_0 \alpha t)^2 + \sin(2 g_0 \alpha t) \mathfrak{Im}[c_g c_e^\ast]$.

If we include also the second term in the expansion Eq.~\eqref{eq:sup_sqrtNapprox}, the probability Eq.~\eqref{eq:sup_peExact} can be calculated in the continuous limit by replacing $c_n$ by a Gaussian distribution and the summation over $n$ by an integral. We obtain
\begin{equation}
    P_e(t) \approx \dfrac{1}{2} \left( 1 + e^{-2 (g_0 t/2)^2} \left( (\abs{c_e}^2-\abs{c_g}^2) \cos(2 g_0 \alpha t) + 2 \mathfrak{Im}[c_g c_e^\ast] \sin(2 g_0 \alpha t) \right) \right) \;. \label{eq:supp_pe2}
\end{equation}
From this result, we notice that the amplitude of the population oscillations decays as $e^{-(t/t_{\text{collapse}})^2}$, where $t_{\text{collapse}}=\sqrt{2}/g_0$ is the collapse time. The origin of this collapse is due to the fact that an atom interacting with a field in a Fock state $\ket{n}$ undergoes Rabi oscillations at a frequency $\sim\sqrt{n}$, and that a field in a coherent state corresponds to a Poissonian distribution of Fock states. The destructive interference between oscillations with different frequency components that dephase results in a decaying signal analogous to the one originating from an inhomogeneously broadened ensemble of atoms.

Equation~\eqref{eq:supp_pe2} describes the collapse of Rabi oscillations, but it does not predict revivals. In fact, it is the discrete spectrum of oscillation frequencies that leads to revivals. Intuitively, this can be seen from the fact that oscillating terms $\cos(g_0\sqrt{n}t)^2 \approx \cos(g_0 n t/\alpha + g_0 \alpha t)$ interfere constructively when $g_0 t/\alpha = 2 \pi q$, with $q = 1,2,...$. This motivates the definition of the revival time as $t_R = 2 \pi \alpha/g_0$. To see this effect concretely, one can for example compute Eq.~\eqref{eq:sup_peExact} numerically by truncating the sum to a suitably large $n_{\text{max}}$.

Here we note that the revivals in the Rabi oscillations are a genuinely quantum feature arising from the discrete spectrum, since if the field were classical, the frequencies would have a continuous spectrum, and such re-phasing could never occur within a finite time. Interestingly, this revival does not tend to a coherent state in the limit of large $\alpha$. Or, in other words, the amplitude of the Rabi oscillations at the revival points never reaches unity. To see this, we can use the results of Ref.~\cite{EberlyPRL80}, where a bound for the revival contrast is presented (Eq.~(6) there). Evaluating this expression at $t_R$ results in a contrast that approaches $B(t_R)=(1+\pi^2)^{-1/4}\simeq0.55$ in the limit of large $\alpha$.

Of particular interest is the phonon state at half the revival time, $t_C \equiv t_R/2 = \pi \alpha  / g_0$, since it takes the form of a cat state. 
To see this, first note that for large $\alpha$ the $c_n$ coefficients vary slowly with $n$ (as they represent a distribution with standard deviation $\sim \alpha$), and therefore one can approximate $c_{n-1}$ and $c_{n+1}$ with $c_n$ in Eq.~\eqref{eq:sup_exactState}.
Moreover, around $t_C$ we can write $g_0 t \sqrt{n+1} \simeq g_0 t \sqrt{n} + \pi/2$ \cite{GeaPRL90}. With these approximations, the state Eq.~\eqref{eq:sup_exactState} simplifies to
\begin{equation}
    \ket{\Psi(t)} \simeq (\ket{g}-i\ket{e}) \sum_{n=0}^\infty \left[  c_g \cos(g_0\sqrt{n} t) - i c_e \sin(g_0 \sqrt{n} t) \right] c_n \ket{n} \;.\label{eq:supp_catTimeState}
\end{equation}
This is a product state, meaning that at $t\approx t_C$ the qubit and the bosonic mode disentangle. Moreover, note that the qubit state at this time does not depend on its initial state at time $t=0$.

\subsubsection{Evolution in the $x$-basis}\label{Sup A.1}
The previous description of the atom-field dynamics in the qubit's $Z$-basis is especially suited for explaining collapse and revivals of the Rabi oscillations. However, to build a better intuition of what happens to the phonon state, it is more instructive to look at the state dynamics in a different basis. For the phonon mode initially in a coherent state $\ket{\alpha}$ with $\alpha \in \mathbb{R}$, we consider the qubit's $X$-basis $\ket{\pm X}=(\ket{e}\pm\ket{g})/\sqrt{2}$. This choice is motivated by the fact that, in the semi-classical limit where the substitution $a\rightarrow\alpha$ is legitimate, the states $\ket{\pm X}$ are eigenstates of $g_0\alpha(\sigma^+ + \sigma^-)$, and thus do not evolve in time. 
In a full quantum description, however, the $\ket{\pm X}$ states also evolve in time according to a rather complex dynamics. The latter can be significantly simplified for $\alpha\gg 1$, where it is possible to write the time evolution of $\ket{\pm X}$ under $H_\mr{JC}$ as \cite{Gea91}
\begin{equation}\label{eq:sup_pmXevol}
    U_\mr{JC}(t)\ket{\pm X}\ket{\alpha} \simeq \dfrac{1}{\sqrt{2}}\left(e^{\mp ig_0 t/2\alpha} \ket{e} \pm \ket{g} \right) \, \ket{\Phi_{\pm}(t)} \;,
\end{equation}
with the phonon states
\begin{equation}\label{eq:sup_fieldStatesApprox}
    \ket{\Phi_{\pm}(t)} \equiv e^{-|\alpha|^2/2} \sum_n \dfrac{\alpha^n}{\sqrt{n!}} e^{\mp ig_0 t\sqrt{n}} \ket{n} =\sum_n c_n e^{\mp ig_0 t\sqrt{n}} \ket{n} \;.
\end{equation}
Eq.~\eqref{eq:sup_pmXevol} can be used for any finite time $t$, and it holds in the sense that the difference between the left and the right hand sides is a state vector whose norm vanishes in the limit $\alpha\rightarrow\infty$.

As we have seen in Eq.~\eqref{eq:sup_sqrtNapprox}, for large $\alpha$ and $t \ll \alpha^2/g_0$, it is possible to approximate the (nonlinear) term $\sqrt{n}$ by a second-order series expansion. When $t \ll \alpha/g_0$, the term quadratic in $n$ can also be neglected, and we can write

\begin{equation}\label{eq:sup_PhiApproxSqrt}
    \ket{\Phi_{\pm}(t)} \simeq e^{\mp ig_0\alpha t/2} \ket{\alpha e^{\mp ig_0t/2\alpha}} \;,
\end{equation}
which shows that for an atom in $\ket{\pm X}$, the evolution of the phonon state is simply a rotation in phase-space with angular velocity $\mp \abs{g_0/2\alpha}$ (see Fig.~1 of the main text). Note that the cat and revival times, $t_C$ and $t_R$, can be directly inferred by setting $g_0t/2\alpha=\pi/2$ or $\pi$, respectively. However, when $t\sim \alpha/g_0$, the term quadratic in $n$ appearing in Eq.~\eqref{eq:sup_sqrtNapprox} cannot be neglected, because the Poissonian distribution of $c_n$ results in a standard deviation of $(n-\alpha^2)$ of order $\alpha$. This implies that for such timescales the nonlinearity of $\sqrt{n}$ plays a significant role in the phase evolution of Eq.~\eqref{eq:sup_fieldStatesApprox}, which results in $\ket{\Phi_{\pm}(t)}$ becoming distorted (and squeezed) coherent states at the cat time $t_C$.

From Eq.~\eqref{eq:sup_pmXevol} it is straightforward to derive the system evolution for any qubit initial state $\ket{\psi_q} = (c_+ \ket{+X} + c_- \ket{-X})$. In particular, for $t_C = \pi \alpha/g_0$ we obtain (cfr. Eq.~\eqref{eq:supp_catTimeState})
\begin{equation}
    U_\mr{JC}(t_{C})\ket{\psi_q}\ket{\alpha} \simeq \dfrac{1}{\sqrt{2}}\left(\ket{g}-i\ket{e}\right) \left( c_+ \ket{\Phi_+(t_{C})} + c_- \ket{\Phi_-(t_{C})} \right) \;.
\end{equation}
From this expression it is easy to see that the qubit state at $t=t_{C}$ is always $\ket{-Y}=(\ket{g}-i\ket{e})/\sqrt{2}$, irrespectively of its initial state $\ket{\psi_q}$. The latter, in fact, gets fully mapped into the phonon state that takes the form of a superposition of \ket{\Phi_{\pm}(t)} with relative complex amplitudes equal to $c_\pm$. 

If for simplicity we consider the approximation Eq.~\eqref{eq:sup_PhiApproxSqrt} and assume $\alpha^2$ to be an even integer, we obtain (omitting a global phase)
\begin{equation}
    U_\mr{JC}(t_{C})\ket{\psi_q}\ket{\alpha} \simeq \dfrac{1}{\sqrt{2}}\left(\ket{g}-i\ket{e}\right) \left( c_+ \ket{-i \alpha} + c_- \ket{i \alpha} \right) \;.
\end{equation}
While for $\ket{\psi_q}=\ket{\pm X}$ the field is simply in a coherent state, for $\ket{\psi_q}$ on the $YZ$-plane ($c_+=1/\sqrt{2}$, $c_-=e^{i\vartheta}/\sqrt{2}$) we obtain for the field the cat state $(\ket{- i \alpha} + e^{i\vartheta} \ket{i \alpha})$. This shows how the phase of the cat state can be controlled through the qubit initial state.

\begin{figure}
\centering
\includegraphics[width=15cm]{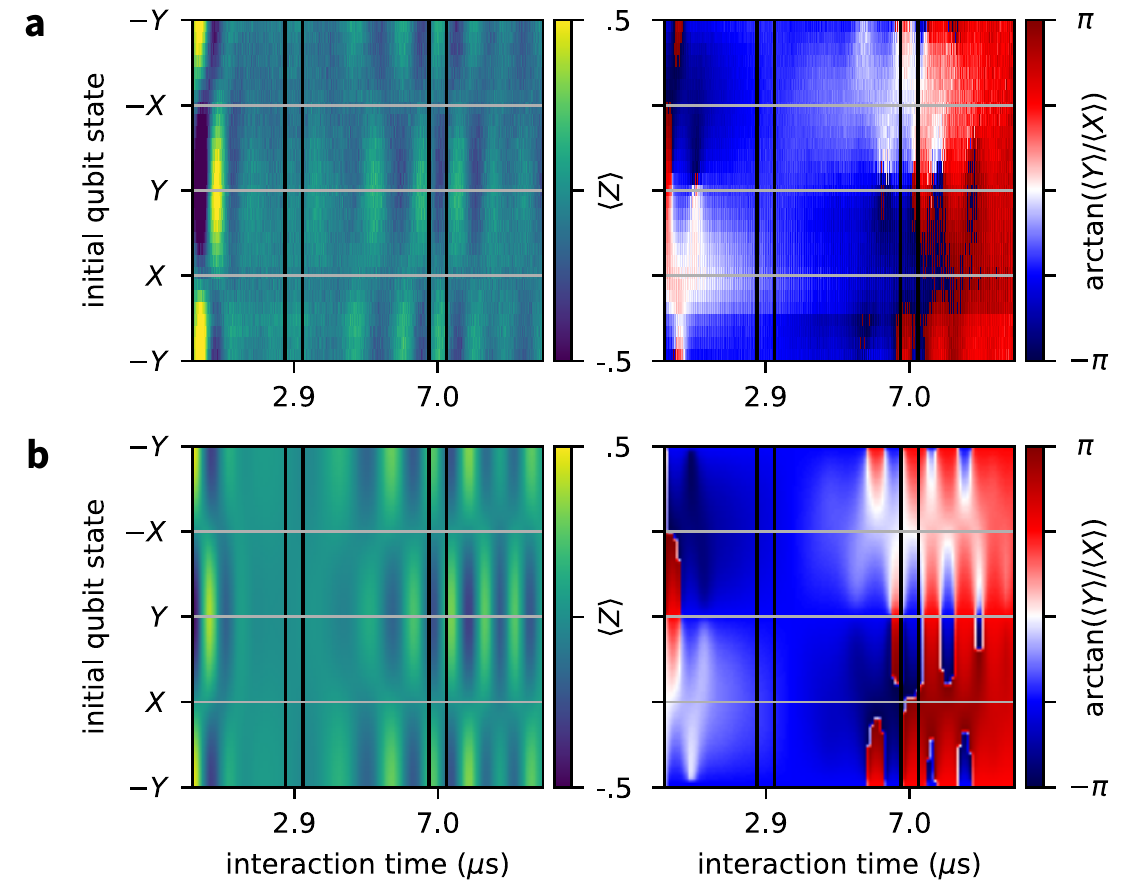}
\caption{\textbf{Collapse and revival for different initial qubit phase.}
\textbf{(a)} Qubit state tomography during collapse and revival when starting with the qubit in $(\ket{g}+e^{i\varphi}\ket{e})/\sqrt{2}$, sweeping $\varphi$ on the y-axis. Left: $\langle Z \rangle$.  When the qubit starts in $\ket{\pm X}$, no collapse and revival is observed. On the other hand, when the qubit starts in $\ket{\pm Y}$, it is observed. Right: Angle between $\langle Y \rangle$ and $\langle X \rangle$. At $t=2.9\,\mu$s, the qubit state is $\ket{-Y}$ for all $\varphi$. At $t=7\,\mu$s, we observe a revival of the qubit state, best visible in $\langle Z \rangle$. \textbf{(b)} Corresponding simulations.
}
\label{fig:suppFig_cnr}
\end{figure}

To demonstrate that the qubit state at $t_{C}$ is in fact $\ket{-Y}$, regardless of the initial state of the qubit, we perform qubit state tomography during the collapse and revival measurement shown in Figure \ref{fig2}b. 
We repeat this measurement for initial qubit states $(\ket{g}+e^{i\varphi}\ket{e})/\sqrt{2}$, varying $\varphi$. The resulting expectation values $\langle Z\rangle$, as well as the resulting qubit phase $\arctan [ \langle Y\rangle / \langle X\rangle ]$, are shown in Fig.\ref{fig:suppFig_cnr}a, from which we make two observations:
First, we see collapse and revival of the qubit population for different initial qubit states, except when the qubit is initially prepared in $\ket{\pm X}$, where the qubit population is stationary. Second, at $t=t_c=2.9\,\mu$s, where we expect a cat state to form in the phonon mode, the qubit state is found to be in $\ket{-Y}$ (blue), regardless of $\varphi$. Fig.\ref{fig:suppFig_cnr}b shows the simulated dynamics of the experiment with the same initial state size and disregarding dissipation, which exhibits good agreement with the measured data. 

\clearpage
\newpage
\subsection{Lattice displacement and effective mass}\label{si:effective mass}
\hfill\\
Analogously to the electromagnetic field in an optical cavity, the strain field in our device can be described by a Laguerre-Gaussian (LG) mode. Considering a beam waist of $w_0 =\unit{27}{\mu m}$ (radius for $1/e$ of the amplitude) and a wavelength of $\lambda = \unit{1.7}{\mu m}$, we obtain a Rayleigh length $z_R = \pi w_0^2 / \lambda \approx \unit{1.4}{mm}$, which is about three times larger than the length $L=\unit{435}{\mu m}$ of the mode itself. For this reason, we approximate the transverse profile of the LG mode as a constant function, which will considerably simplify the following calculations. This function takes the form
\begin{equation}\label{eq:sup_LG}
    LG_{pl}(r,\phi) = \sqrt{\dfrac{2 p!}{\pi(p+\abs{l})!}} \left(\dfrac{r \sqrt{2}}{w_0}\right)^{\abs{l}} e^{-(r/w_0)^2} L_p^{\abs{l}}\left(\dfrac{2 r^2}{w_0^2}\right) e^{-i l \phi} \;,
\end{equation}
where the integer $p\geq 0$ is the radial index, $l \in \mathbb{Z}$ is the azimuthal index, and $L_p^l$ are the generalized Laguerre polynomials. These modes are normalized so that $\int_0^{2 \pi} d\phi \int_{0}^{\infty} dr \, r \abs{LG_{pl}(r,\phi)}^2 = w_0^2$. 

The strain field can now be written as
\begin{equation}\label{eq:supp_strain}
    s_{plm}(r,\phi,z) = S \, LG_{pl}(r,\phi) \, \sin\left( \dfrac{m \pi}{L} z\right) \;,
\end{equation}
where $S$ is a normalization constant, and $m \pi/ L = k = 2\pi / \lambda$, with the integer $m\geq 1$ being the longitudinal index. The normalization is such that
\begin{align}
    U &= \dfrac{c_{33}}{2} \int_V dV \abs{s_{plm}(r,\phi,z)}^2 \\
    &= \dfrac{c_{33}}{2} S^2 w_0^2 \dfrac{L}{2} \label{eq:supp_S0}
\end{align}
is the potential energy stored in the deformation of the crystal, with $c_{33}$ the material stiffness tensor component. In this sense, $S$ depends on the system's state. If we set $U$ to be the energy of a phonon $\hbar\omega_p$, Eq.~\eqref{eq:supp_S0} defines $S=S_0\equiv\sqrt{4\hbar\omega_p/(L w_0^2c_{33})}$ as the strain per phonon.

To find the effective mass of the mechanical mode of interest, it is convenient to compare it to a one-dimensional harmonic oscillator of mass $\Meff$ and potential energy
\begin{equation}\label{eq:supp_HOeff}
    U = \dfrac{1}{2} \Meff \omega_p^2 \xeff^2 \;,
\end{equation}
where $\xeff$ is the effective oscillation amplitude.
This expression alone is not sufficient to define  $\Meff$ and $\xeff$ unambiguously, as it constraints only their product. Equating Eq.~\eqref{eq:supp_HOeff} and Eq.~\eqref{eq:supp_S0}, we get
\begin{align}
    \Meff &= \dfrac{c_{33}}{2}\dfrac{S^2 w_0^2}{2 \omega_p^2 \xeff^2} L \\
    &= \left( \dfrac{S^2 L^2}{2 \pi^3 m^2 \xeff^2}\right) \rho \pi w_0^2 L \label{eq:supp_Meff}
\end{align}
where in going to the last line we used the relations $\omega_p=2\pi c/ \lambda$, $\lambda=2 L/m$ and $c=\sqrt{c_{33}/\rho}$. 

The term in parenthesis of Eq.~\eqref{eq:supp_Meff} is a rescaling factor of the mass $M_0 \equiv \rho \pi w_0^2 L=\unit{4.0}{\mu g}$. Computing it requires an estimate of $\xeff$. 
Note here that $M_0$ is the mass of a cylinder with radius $w_0$ and the length $L$ and density $\rho$, which is the naive approximation for the volume of the phonon mode. 
To obtain an estimate of $\xeff$, note that Eq.~\eqref{eq:supp_strain} describes the strain in the material, which we assumed to be mostly in the longitudinal (i.e. $z$) direction. This is defined as $s\equiv s_{zz}=\partial u_z(x,y,z)/\partial z$, where $u_z(x,y,z)$ is the $z$ component of the displacement field $\vec{u}(x,y,z)=( 0,0,u_z(x,y,z) )$. Therefore, $u_z$ can be found by a straightforward integration of Eq.~\eqref{eq:supp_strain} to be
\begin{equation}
    u_z(r,\phi,z) = - \dfrac{L}{m\pi} S \, LG_{pl}(r,\phi) \, \cos\left(\dfrac{m \pi z}{L}\right)  \;.
\end{equation}

The maximum displacement can be easily found for $l=0$, for which the maximum of Eq.~\eqref{eq:sup_LG} is $LG_{p0}(0,0)=\sqrt{2/\pi}\approx 0.8$. This gives $u_z^{\text{max}} = \frac{L}{m \pi}\sqrt{\frac{2}{\pi}} S$. On the other hand, a root-mean-square (RMS) value can be found by defining an integration area $A_k = \pi R^2$, such that for $R=2 w_0$ and the values $p=0,1$ of typical interest we find
\begin{equation}
\sqrt{A_k^{-1} \int_0^{2 \pi} d\phi \int_{0}^{R} dr \, r \abs{LG_{p0}(r,\phi)}^2} \approx 0.28 \;.   
\end{equation}
Together with the factor $1/\sqrt{2}$ coming from the RMS of the cosine, we have $u_z^{\text{RMS}} = \frac{L}{m \pi}\frac{0.28}{\sqrt{2}} S$.

Choosing $\xeff=u_z^{\text{max}}$ results in $\Meff=M_0/4 \approx \unit{1.0}{\mu g}$, while for $\xeff=u_z^{\text{RMS}}$, we obtain $\Meff \approx 4.1 M_0 \approx \unit{16.2}{\mu g}$.

To find the displacement of the lattice for a coherent state, we use Eq.~\eqref{eq:supp_HOeff} and the relation $U=\hbar\omega_p(\alpha^2+1/2)$ to write
\begin{align}
    \xeff(\alpha) &= \sqrt{1+2\alpha^2} \; \sqrt{\dfrac{\hbar}{\Meff \omega_p}} \\
    &= \sqrt{2(1+2\alpha^2)} \; x_{\text{ZPF}} \;,
\end{align}
where $x_{\text{ZPF}} \equiv \sqrt{\hbar / 2 \Meff \omega_p}$ is the zero point fluctuation.
Here, $\xeff^{\text{RMS}}$ or $\xeff^{\text{max}}$ is obtained by choosing the corresponding $\Meff$.
For example, a cat state with $\alpha=1.61$ would correspond to a mass of $\Meff = \unit{1.0}{\mu g}$ delocalized over $2 \xeff^\text{max}(1.61) = \unit{8.4 \cdot 10^{-18}}{m}$, or a mass of $\Meff = \unit{16.2}{\mu g}$ delocalized over $2 \xeff^\text{RMS}(1.61) = \unit{2.1 \cdot 10^{-18}}{m}$.

\clearpage
\subsection{Parity measurement calibration}\label{si: parity_norm}
\hfill\\
\begin{figure}
\centering
\includegraphics[width=7cm]{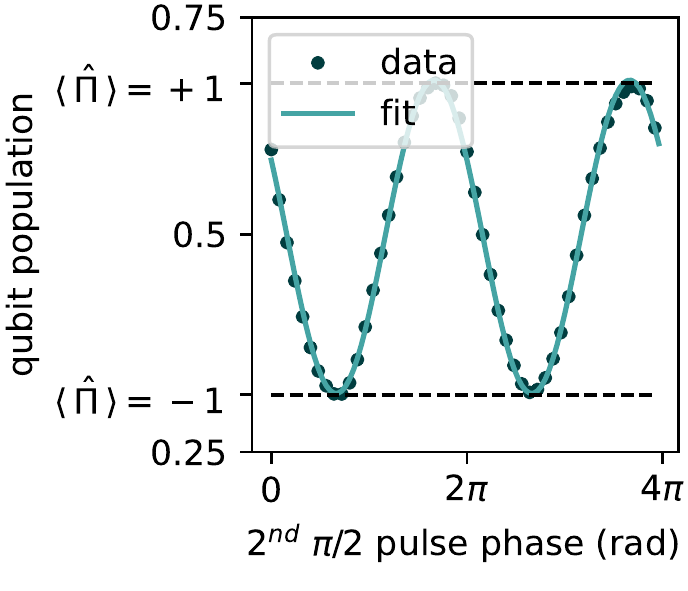}
\caption{\textbf{Parity measurement calibration curve.} The two dashed black lines indicate values of maxima and minima of the parity values for a Fock $\ket{0}$ state. These two values are then used as the $\pm 1$ parity values to normalize subsequent Wigner tomography measurements.}
\label{fig:suppParityNorm}
\end{figure}
To perform the phonon parity measurement needed for Wigner tomography, we follow the procedure developed in earlier work\cite{vonLupke22}. This consists of a Ramsey-type qubit experiment, where we let the qubit dispersively interact with the phonon mode for a carefully calibrated interaction time. After this interaction, the phonon state parity has mapped onto the qubit state due to the phonon-state-dependent dispersive frequency shift of the qubit. Ideally, even (odd) phonon numbers map to the $\ket{e}$ ($\ket{g}$) state of the qubit. Measuring the qubit state then corresponds to a measurement of the phonon parity. 

However, because of finite qubit coherence, the contrast of the parity measurement is reduced. For example, the parity measurement of an undisplaced Fock state $\ket{0}$ results in a probability $P(e\vert 0)<1$, which translates to a parity value $\langle\hat{\Pi}\rangle < 1$. To correct for this measurement error, we normalize the measured parity values using the contrast of the Ramsey parity measurement with the phonon in $\ket{0}$, obtained by sweeping the phase of the second $\pi/2$ pulse. An example for an interaction time of $6.7\,\mu$s can be seen in Fig.\ref{fig:suppParityNorm}. 
$P(e\vert 0)$ oscillates between the maximum and minimum values possible at the given interaction time. Fitting this oscillation to a cosine model, we extract the oscillation amplitude and use it to normalize all parity measurements so that the maximum and minimum readout values correspond to a parity of $+1$ and $-1$, respectively.

\subsection{Resonant phonon number measurement}\label{si:resonant measurement}
\begin{figure}
    \centering
    \includegraphics[width=12cm]{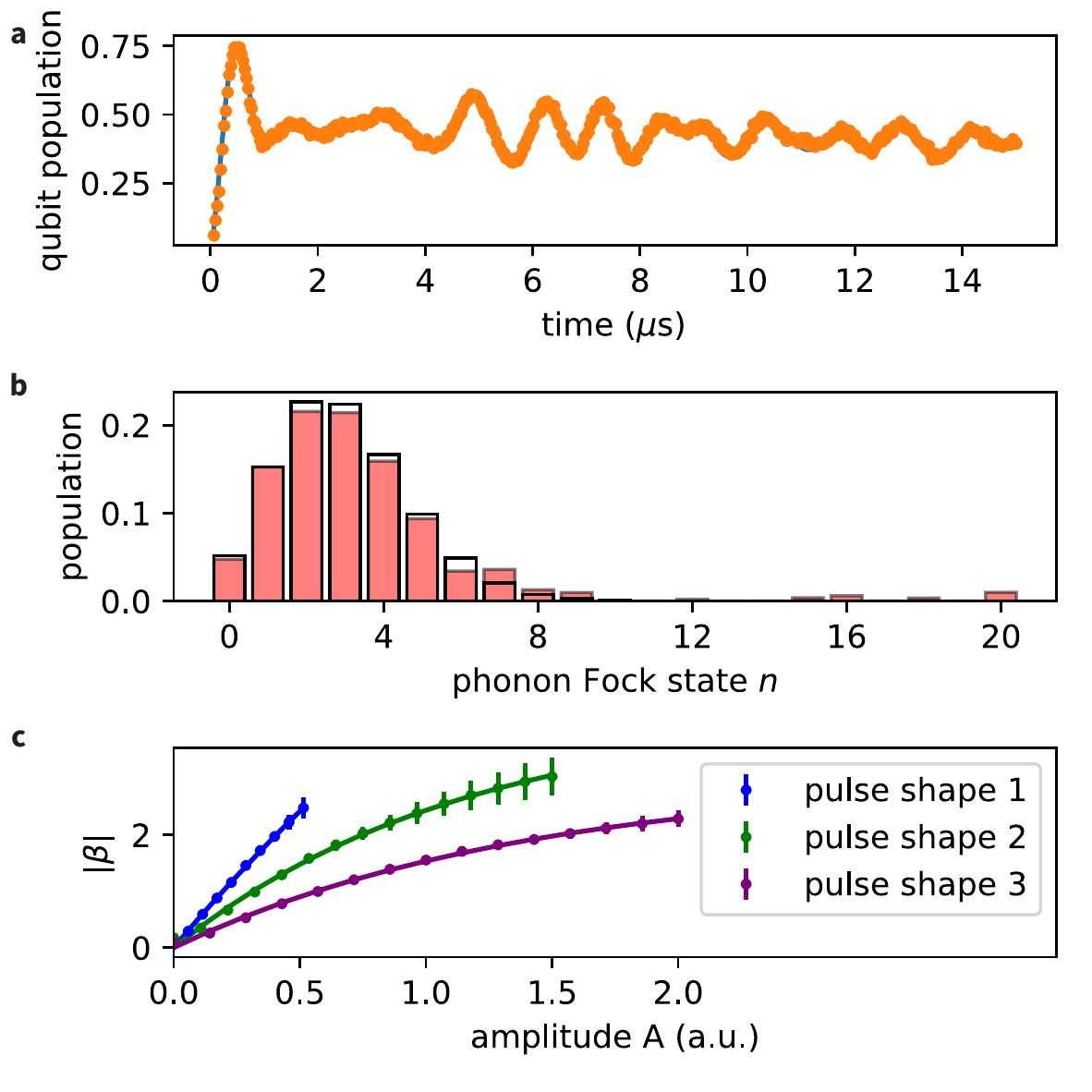}
    \caption{\textbf{Phonon displacement drive calibration.} 
    \textbf{(a)} An example of the qubit dynamics during resonant interaction with the phonon mode. Orange dots are experiment data, and the blue curve is the fitting result.
    \textbf{(b)} Distribution of Fock state populations extracted from the measurement of \textbf{a}. The black boxes are the Poisson distribution of the fitted coherent state.
    \textbf{(c)} Relation between the drive amplitude and generated coherent state $\abs{\beta}$ for three different pulse shapes. 
    }
    \label{fig:resonance phonon number}
\end{figure}

\hfill\\
Due to limitations in our setup, the magnitude $\abs{\beta}$ of the coherent state we generate may not scale linearly with the drive amplitude $A$. We find that $\abs{\beta}$ saturates for larger $A$, when the drive strength on the qubit is close to the detuning between qubit and phonon. Therefore, we calibrate the generated $\abs{\beta}$ by measuring the phonon Fock state populations for different $A$, using the method described in earlier works\cite{Chu2018}. Figure \ref{fig:resonance phonon number}a is an example time trace showing the qubit population as it interacts resonantly with the phonon initialized in a coherent state. We first perform a fit of this data to extract the phonon Fock state populations and then fit these populations with a Poissonian distribution (Fig. \ref{fig:resonance phonon number}b). In Fig.\ref{fig:resonance phonon number}c, we show the extracted $\abs{\beta}$ as a function of $A$ for three different pulse shapes, which we fit to a phenomenological function of the form $\abs{\beta}=C\left(\text{exp}(A/B)-1\right)$. The pulse shapes are all Gaussian-square pulses, and the three examples have different lengths and rise times, with the amplitude $A$ as an overall scaling factor. When performing Wigner tomography, we use a linear grid of drive amplitudes for convenience. In post-processing, we then use the fitted dependence of $\abs{\beta}$ on $A$ to plot the parity values as a function of the actual coherent displacement amplitudes.

\subsection{Post-processing and fitting of Wigner functions}
\begin{figure}[h!]
    \centering
    \includegraphics[width=12cm]{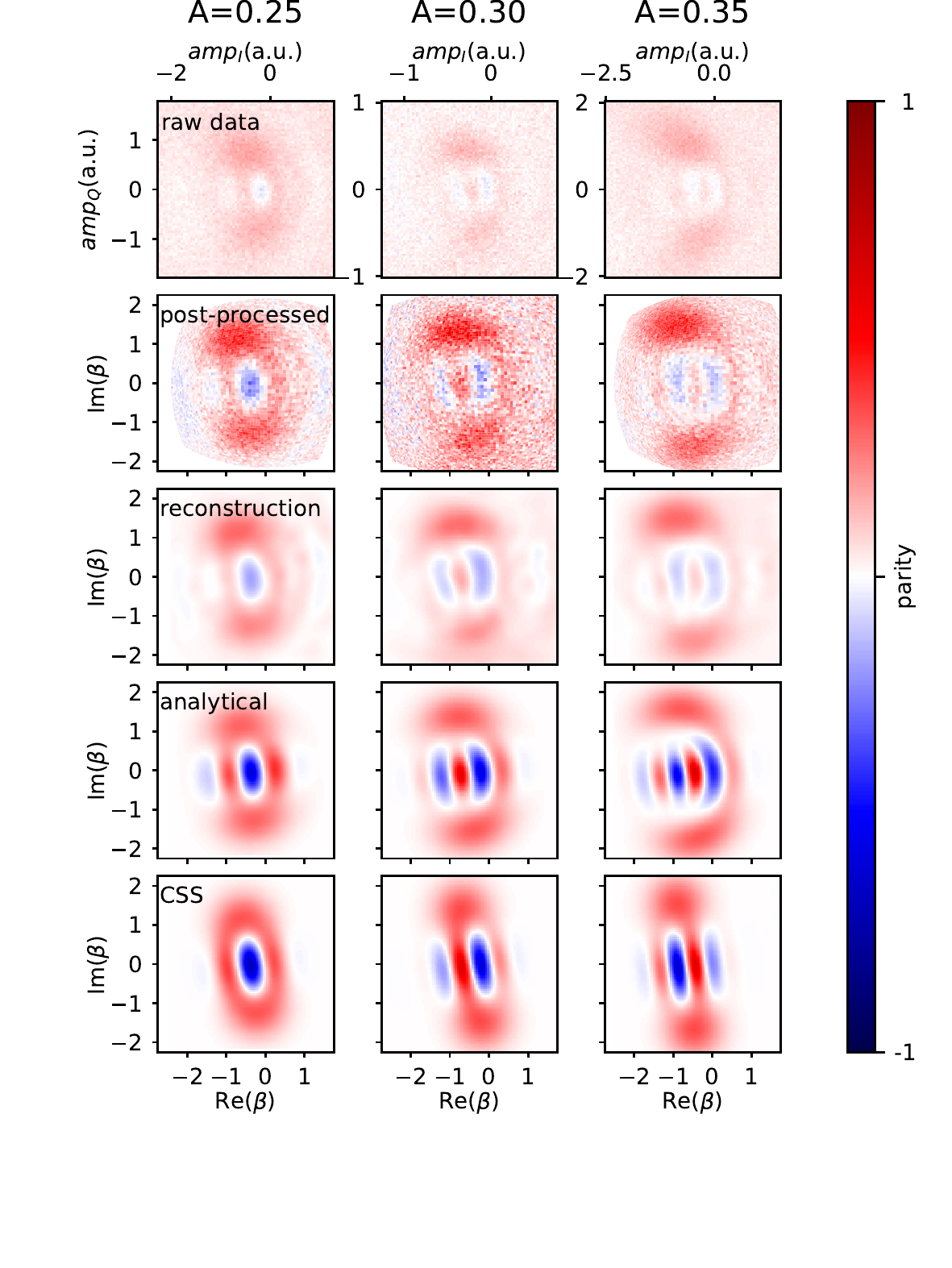}
    \caption{\textbf{Post-processing and analysis of Wigner functions.} Each column corresponds to one of the three initial displacement amplitudes $A$ used to prepare the cat states shown in the main text. The axes for the first row are the $I$ and $Q$ quadrature amplitudes of the complex drives used in the Wigner tomography sequence\cite{Chu2020, vonLupke22}. These are then converted into the real and imaginary parts of the actual displacement amplitude using the procedure in section \ref{si:resonant measurement}. 
    }
    \label{fig:wigner_table}
\end{figure}

\hfill\\
In Fig.\ref{fig:wigner_table}, we show an overview of how the measured Wigner functions are post-processed and fitted to different cat state models. We start with the raw data in the first row, which is plotted as a function of the complex drive amplitude. We then normalize the measured parity values using the procedure described in section \ref{si: parity_norm} and perform a nonlinear scaling of the axes as described in section \ref{si:resonant measurement}. The resulting post-processed data is shown in the second row of Fig.\ref{fig:wigner_table}, where the Wigner function is now plotted as a function of the actual complex displacement amplitude. Rows three to five then show the result of maximum-likelihood reconstructions of the physical states from the post-processed data, as well as fits to the analytical and CSS cat state models, as described in section \ref{si:reconstruction} and in the main text. 

\clearpage
\newpage
\subsection{State reconstruction and fidelity}\label{si:reconstruction}
\hfill\\
\begin{figure}
\centering
\includegraphics[width=15cm]{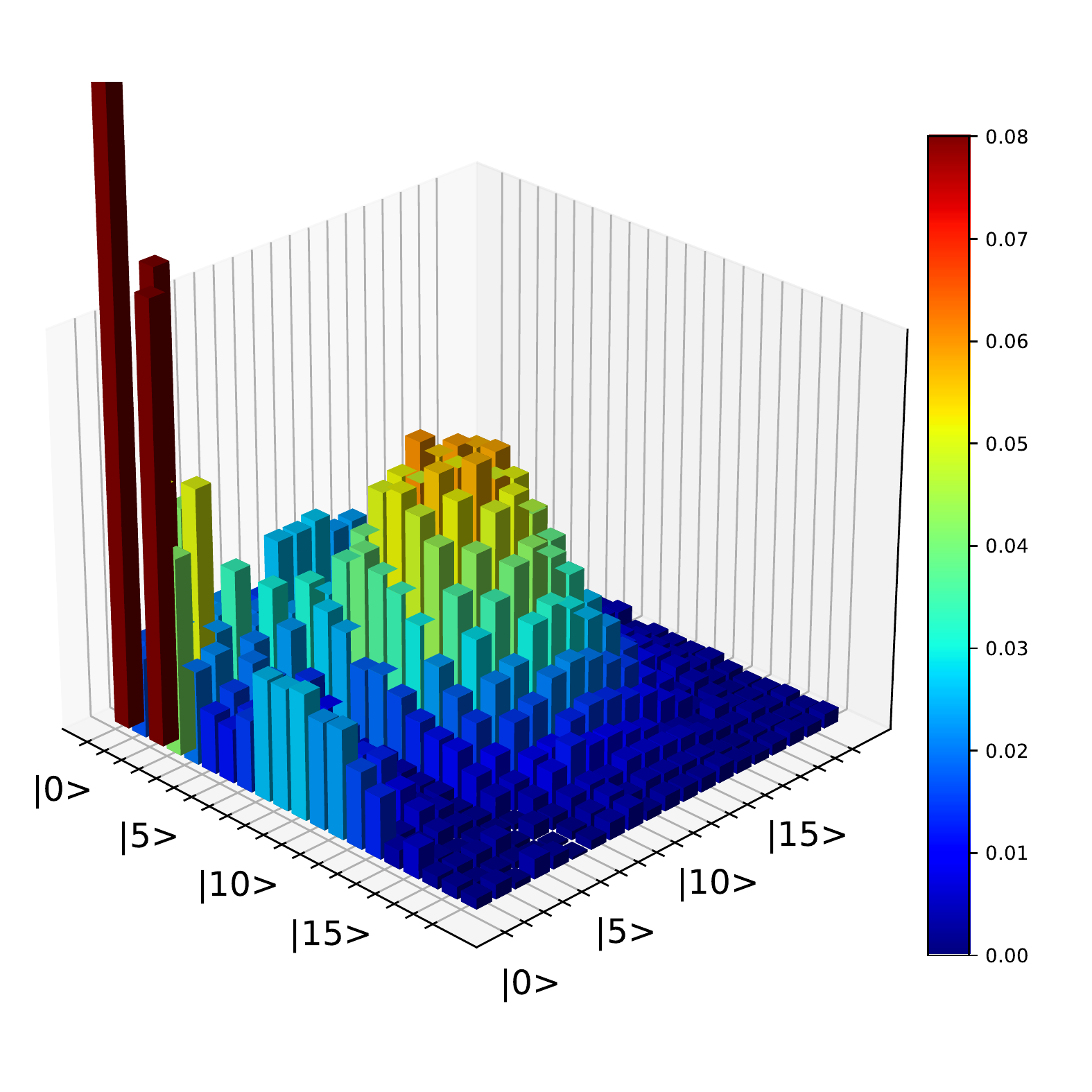}
\caption{\textbf{Reconstructed density matrix for the cat state with A=0.35.} The height and color of the bars both indicate the absolute value of the corresponding density matrix elements.
\label{fig:DM} }
\end{figure}

Reconstruction of the density matrix from the measured Wigner function is performed using a Maximum Likelihood Estimate, as described in\cite{Chou2018}. As an example, Fig.~\ref{fig:DM} shows the absolute value of the displaced reconstructed density matrix of the $A$=0.35 cat state from the main text. The additional displacement shifts one of the coherent state components to the center of phase space.  The two peaks on the diagonal represent the coherent state components. In turn, the features along the coordinate axes indicate interference between the coherent state components, corresponding to fringes of alternating parity in phase space. 

We then calculate the fidelity of the reconstructed state with respect to two different target states. Here we define the fidelity between two states $\rho$ and $\sigma$ as
\begin{equation}
    \mathcal{F}(\rho, \sigma) = \text{Tr}\left(\sqrt{\sqrt{\rho} \sigma \sqrt{\rho}}\right) \;.
\end{equation}

The first target state we consider is the phonon state resulting from the JC interaction between the qubit and phonon modes, as described by Eq.~\ref{eq:sup_exactDM} (up to rotations). To most accurately model the physical state produced in our system, we fix the initial qubit state (described by $c_g$ and $c_e$) and JC evolution time $t = t_C$ of each fitted state to match those of the corresponding experiment. We also allow for  a rotation $\theta$ in phase space, which could result from a slight detuning of the phonon drive frequency. This leaves a target state of the form
\begin{equation}
    \rho(t_C) = R(\theta)\rho'(t_C) R^{\dagger}(\theta)\;,
\end{equation}
where $\rho'(t_C)$ is given by Eq.~\ref{eq:sup_exactDM} with $t=t_C$ and $R(\theta) = \text{exp}(-i \theta a^{\dagger}a)$.
The parameters $\alpha$ and $\theta$ are optimized numerically in order to maximize the fitted state's fidelity to the reconstructed experimental state. We quote the fitted $\alpha$ values as $\alpha_{\text{fit}}$ in the main text.

The second target state is a coherent state superposition of the form
\begin{equation}
    \ket{\text{C}} = \mathcal{N} \left( \ket{\alpha_1} + e^{i\vartheta} \ket{\alpha_2} \right)\label{eq:css_SI} \;,
\end{equation}
where $\mathcal{N}$ is an appropriate normalization constant. Again, the parameters $\ket{\alpha_{1,2}}$ and $\vartheta$ are optimized numerically in order to maximize the fidelity. From these, we obtain the CSS cat state size as half of the distance between the two coherent states, $D=\vert \alpha_1 - \alpha_2 \vert$ \cite{brune1996observing}. 

\begin{figure}
\centering
\includegraphics[width=6cm]{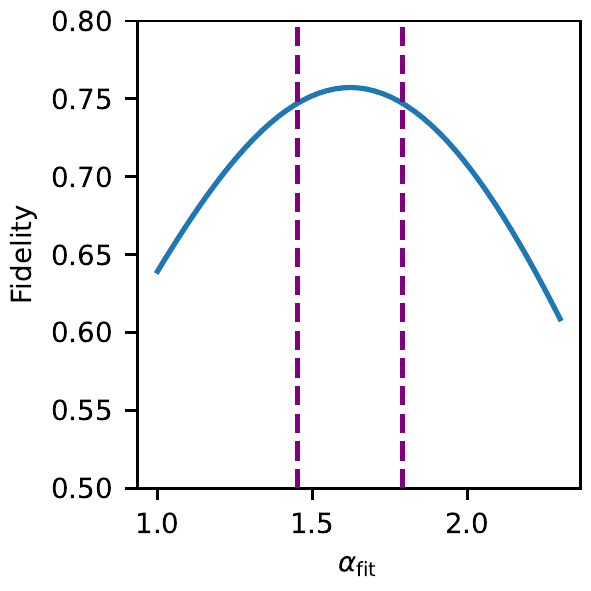}
\caption{\textbf{Fidelity sensitivity to $\alpha_{\text{fit}}$.}
Fitted fidelity of the analytical state to the cat state generated with $A=0.35$. In the fit, we constrain the initial coherent state size $\alpha_\text{fit}$. The two purple dashed lines are the positions where the fitted fidelity is 1\% lower than the highest fidelity. 
\label{fig:suppFidelitySensitivity} }
\end{figure}

To determine how sensitive the fidelities resulting from our fitting procedure are to different values of $\alpha_{\text{fit}}$ and $D$, we repeat the fits, but rather than optimizing $\alpha_{\text{fit}}$ or $D$, we keep them fixed in the fit and sweep across a range of values. Fig.~\ref{fig:suppFidelitySensitivity} shows an example for $\alpha_{\text{fit}}$, from which we then calculate the alpha value where the fidelity is 1\% lower than the highest fidelity and use this range as the error bar in Fig.~\ref{fig3}b.

\clearpage
\newpage
\subsection{Cat state decoherence}\label{si:cat_decoherence}
\hfill\\
We consider here a bosonic mode subject to relaxation at a rate $\kappa$. For an initial coherent state $\ket{\alpha}$, this relaxation process results in the time-dependent state $\ket{\alpha(t)}= \ket{\alpha e^{-\kappa t/2}}$. In Ref.~\cite{BrunePRA92,HarocheBook}, it is shown that the Wigner function of a CSS state evolves in time as
\begin{equation}\label{eq:supp_Wdeccat}
    W(\beta) = \dfrac{1}{\pi(1+e^{-\vert \alpha\vert^2})} \left( e^{-2\vert\beta-\alpha\epsilon\vert^2} + e^{-2\vert\beta+\alpha\epsilon\vert^2} + 2 e^{-2\vert\beta\vert^2}e^{-2\vert\alpha\vert^2 (1-\epsilon^2)} \cos(4 \text{Im}(\beta)\alpha \epsilon) \right) \;,
\end{equation}
where $\alpha$ is the amplitude of the coherent states in the superposition, and $\epsilon=e^{-\kappa t/2}$ parameterizes the relaxation process.

The two exponential terms $e^{-2\vert\beta\pm\alpha\epsilon\vert^2}$ represent the two coherent state components relaxing towards the origin with a timescale $2/\kappa=2 T_1^{ph}$. On the other hand, the amplitude of the cosine term, representing the coherence of the quantum superposition, decays much more rapidly as it includes the factor
\begin{equation}\label{eq:supp_xi}
    \xi \equiv e^{-2\vert\alpha\vert^2 (1-\epsilon^2)} \;.
\end{equation}

To see this effect, it is necessary to access a quantity which is able to reflect this decay of coherence. Here, we show that the Wigner function negativity can be taken as a faithful quantifier for the coherence. The negativity is defined as \cite{Kenfack04}
\begin{equation}\label{eq:supp_WNegDef}
    \delta \equiv \int d^2\beta \Big( \vert W(\beta)\vert - W(\beta) \Big) \;,
\end{equation}
and it is an indicator of non-classicality that has been related to a number of quantum information tasks and entanglement measures \cite{MariPRL12,WalschaersPRL17,AlbarelliPRA18,TanPRL20}.

For the CSS state in Eq.~\eqref{eq:supp_Wdeccat}, the negativity can be computed analytically for large $\alpha$ by taking into account that the overlap between the three exponential terms is negligible. These correspond to the two coherent states and to the interference fringes in between them, which for large $\alpha$ are well separated in phase space. This allows us to approximate $\vert W(\beta)\vert$ by the sum of the moduli of each exponential term, giving (using $\int d^2\beta W(\beta) = 1$)
\begin{align}
    \delta_{\text{cat}}(\epsilon) &\approx \dfrac{1}{\pi(1+e^{-\vert \alpha\vert^2})} \int d^2\beta \left( \vert e^{-2\vert\beta-\alpha\epsilon\vert^2} \vert + \vert e^{-2\vert\beta+\alpha\epsilon\vert^2} \vert + \vert 2 e^{-2\vert\beta\vert^2}e^{-2\vert\alpha\vert^2 (1-\epsilon^2)} \cos(4 \text{Im}(\beta)\alpha \epsilon) \vert \right) - 1 \notag\\
    &= \dfrac{1}{2}(1+\tanh(\alpha^2)) + I_{\text{decay}}(\epsilon) - 1 \;.
    \label{eq:supp_catNegDec}
\end{align}
Here we introduced the term
\begin{equation}\label{eq:Idecay}
    I_{\text{decay}}(\epsilon) = \dfrac{\sqrt{2} e^{2\alpha^2 \epsilon^2}}{\sqrt{\pi}(1+e^{2\vert \alpha\vert^2})} \int d\beta_i  e^{-2\beta_i^2} \vert \cos(4 \beta_i \alpha \epsilon) \vert \;,
\end{equation}
with $\beta_i$ as the imaginary part of $\beta$, which is responsible for the negativity decay. The time constant for this process can be estimated by considering the lower bound $\cos(x)^2 \leq \vert\cos(x)\vert$, which allows us to approximate the integral in $I_{\text{decay}}$ with
\begin{equation}
    \int d\beta_i  e^{-2\beta_i^2} \cos(4 \beta_i \alpha \epsilon)^2 = \sqrt{\dfrac{\pi}{8}} \left( 1 + e^{-8 \alpha^2 \epsilon^2} \right) \;.
\end{equation}
For large $\alpha$ and short times, this expression can be used to find
\begin{equation}\label{supp:decrate2}
    I_{\text{decay}}(\epsilon) \approx \dfrac{1}{4} (1 + \tanh(\alpha^2) ) e^{-2 t \alpha^2 \kappa} \;,
\end{equation}
which shows that the cat state negativity decays at a rate $2 \alpha^2 \kappa$, corresponding to a time constant $\tau_\text{cat} = T_1^{ph} / (2 \alpha^2)$. We note that, by looking at Eq.~\eqref{eq:Idecay}, it's clear that the same decay time constant should apply to the negativity of a 1D slice of the Wigner function perpendicular to the fringes at $\text{Im}(\beta)=0$ if $\alpha$ is purely imaginary, which corresponds to the measured data in Figure 4 of the main text.

\begin{figure}
\centering
\includegraphics[width=10cm]{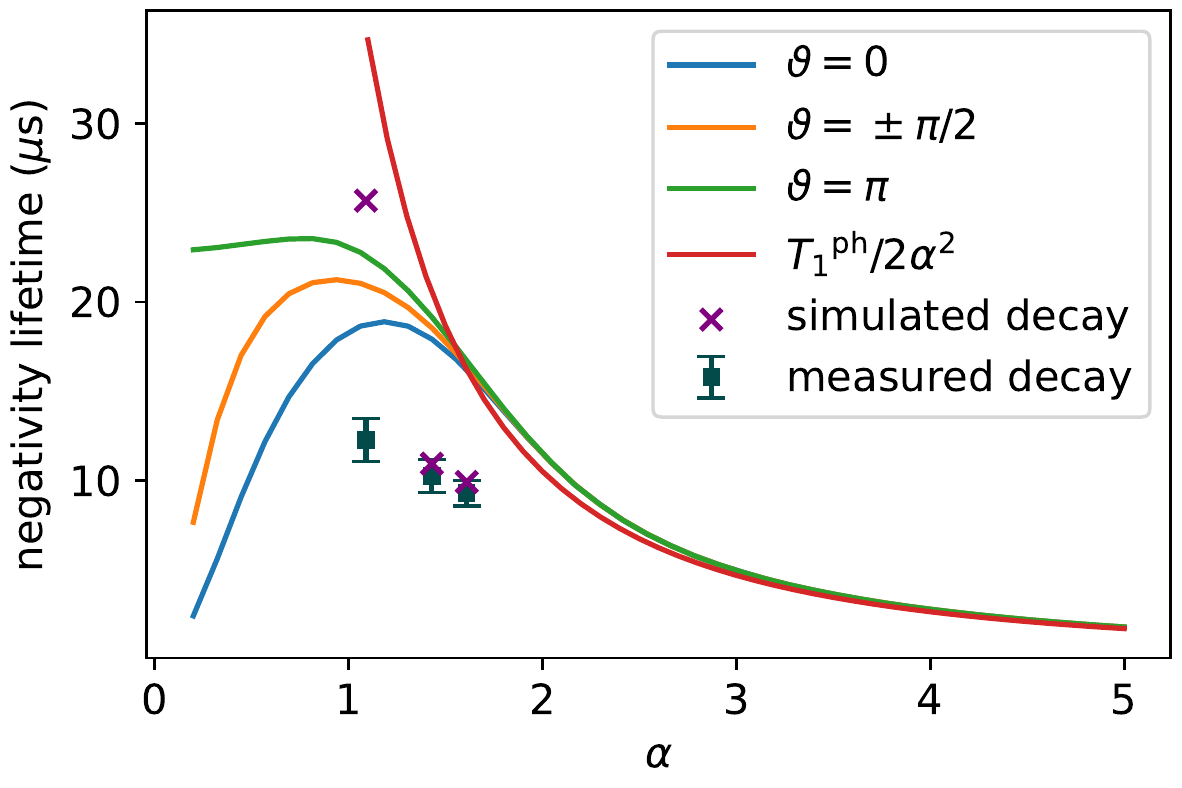}
\caption{\textbf{Cat decay rate predictions and measurements.} Wigner negativity decay rate for CSS with different phases $\vartheta$ in Eq.~\eqref{eq: ideal_cat} (blue, orange, green lines), compared to the analytical decay rate obtained in the limit of large $\alpha$ (red line). Purple crosses are the decay times obtained from a simulation of the full experiment as described in section G.2, black squares are the decay times of the measured states shown in Fig.4c of the main text.
\label{fig:suppFigCatPhaseDecays} }
\end{figure}

In deriving Eq.~\eqref{supp:decrate2}, we assumed the coherent state amplitude $\alpha$ to be large. In order to check the validity range of this approximation, we compare $\tau_\text{cat}$ to the decay rate obtained by numerically computing the negativity of Eq.~\eqref{eq:supp_Wdeccat} for varying $\alpha$. The results are shown in Fig.~\ref{fig:suppFigCatPhaseDecays}, where we can see good agreement for $\alpha \gtrsim 2$. However, for smaller values of $\alpha$ we see a significant discrepancy, as we expect from the fact that the approximation used to obtain Eq.~\eqref{eq:supp_catNegDec} becomes inaccurate. Moreover, when $\alpha$ is small, we observe a decay speed that is also dependent on the phase $\vartheta$ of the coherent states superposition Eq.~\eqref{eq: ideal_cat}.

\subsubsection{Decoherence Measurements}\label{si:decoherence_measurements}
\begin{figure}
\centering
\includegraphics[width=15cm]{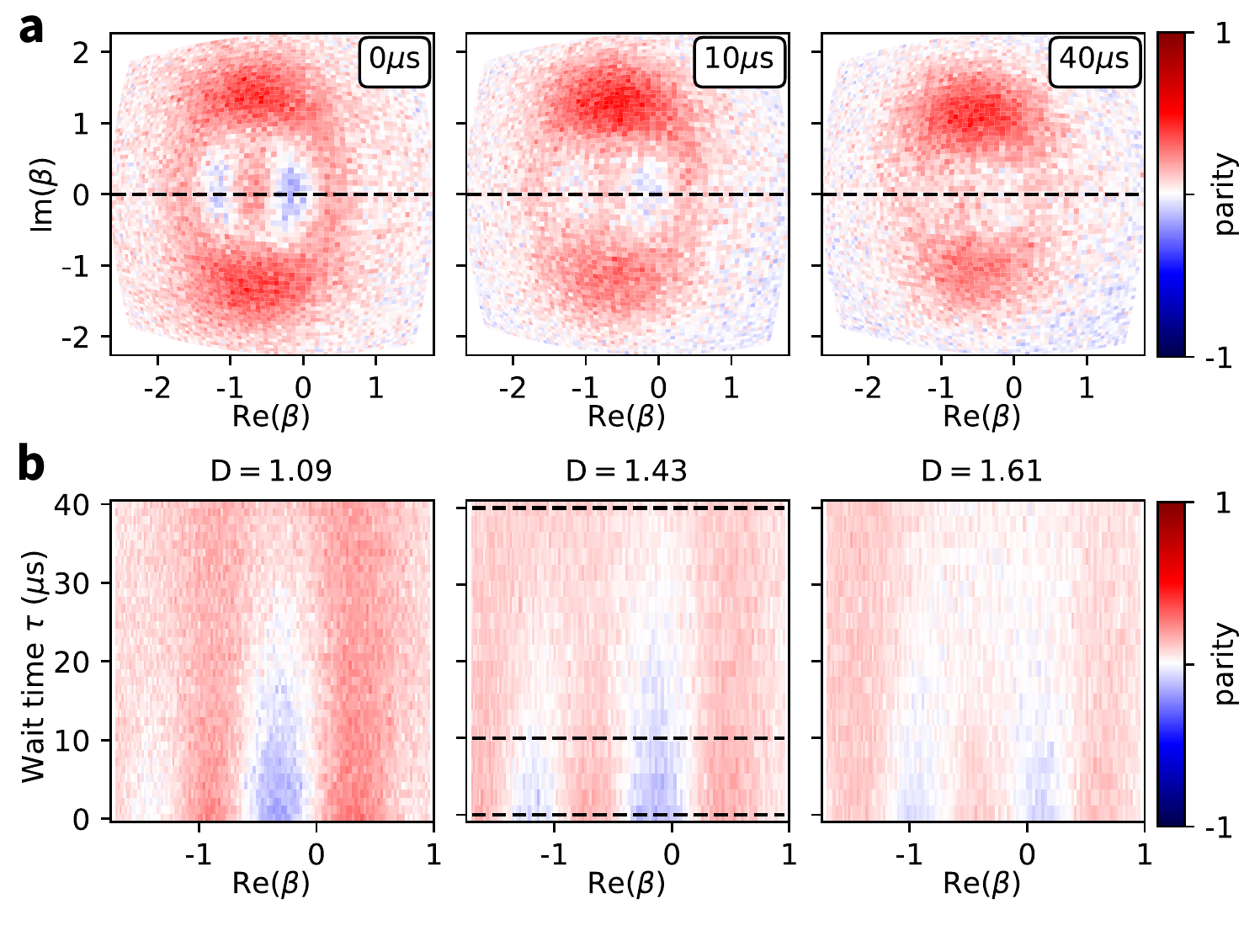}
\caption{\textbf{Wigner negativity decay measurements.} \textbf{(a)} Wigner tomographies of the cat state with size $D=1.43$ for three different wait times. \textbf{(b)} 1D crosscuts vs. wait time for all three measured cat states with size $D$ as indicated in the plot titles. Black dashed lines in \textbf{(a)} and \textbf{(b)} indicate correponding crosscuts.}
\label{fig:suppFigS8}
\end{figure}

In order to characterize the decay of quantum features in the created cat states, we extract the state negativity (see Eq.~\eqref{eq:supp_WNegDef}) after different wait times $\tau$ between state creation and tomography. To illustrate the decaying state in phase space, we perform full Wigner tomography on the cat state of size $D=1.43$ at the delay times $\tau = 0, 10, 40\,\mu$s. The resulting data is presented in Fig.~\ref{fig:suppFigS8}a. While the negative parity regions of the interference fringes are decaying at a fast rate of $\approx 10\,\mu$s, the decay of the coherent state components towards the vacuum state happen at the slower timescale of $2 T_1^{ph}$, as shown in the previous section. 

Extracting the negativity of a state with high enough accuracy requires increasing the number of averages by a factor of $5$ compared to the presented states in Fig.~2 and 3 of the main text. Additionally, the number of delay times $\tau$ between state creation and tomography needs to be large enough to extract fitting parameters with reasonable errorbars. In order to fulfill these requirements efficiently, we choose to measure the decaying states along a 1D crosscut perpendicular to the interference fringes, around the location of largest contrast between positive and negative parity fringes (see Fig.~\ref{fig:suppFigS8}a). 

We plot these crosscuts of the Wigner function for all three cat state sizes and $\tau$ ranging from from $0$ to $40\,\mu$s in Fig.~\ref{fig:suppFigS8}b. For each crosscut, the negativity is calculated according to Eq.~\eqref{eq:supp_WNegDef}. One can clearly observe the faster decay rate with increasing state size, confirmed by the analysis presented in Fig. 4 of the main text.    

\subsubsection{Decoherence simulations for analytical states}
As we have seen from Fig.~\ref{fig:suppFigCatPhaseDecays}, for small values of $\alpha$ we expect a discrepancy between the theoretical and the measured $\tau_\text{cat}$. For this reason, in order to obtain a more accurate prediction of the cat decay time, we perform a Master equation simulation of the full experiment and compare it to the decay of the CSS states shown in Fig.~\ref{fig:suppFigCatPhaseDecays}.

In order to reproduce the state decay as accurately as possible, we first run a full master equation simulation of the cat state preparation, starting from an initial coherent state with same amplitude as in the experiment. After having verified that the simulated states agree well with the corresponding measured cat states at $t_C$, the simulated states are evolved freely for a variable time $\tau_{\text{sim}}$ under Eq.~$\eqref{eq:HJC}$ with the qubit far detuned from the phonon mode. Finally, the states' Wigner negativity is extracted for each $\tau_{\text{sim}}$ following the same procedure as described in section G1. These simulated negativities are fitted to the same decaying exponential model that we also apply to the measured data, allowing us to extract the desired time constant. 

The results obtained for three cat states of the same size as the one we measured can be seen as purple crosses in Fig.\ref{fig:suppFigCatPhaseDecays}. The simulated decay rates are $24.68, 12.34$, and $10.52\,\mu$s, for the state sizes $D=1.09, 1.43$, and $1.61$ respectively. The data shown in Fig. 4c of the main text is also included for comparison. The deviation of the measured value for the smallest state size from the simulated value can be attributed to a slight positive background offset of the measured Wigner function, which results in the negativity values being set to zero in the tail of the decay.

\clearpage
\newpage
\bibliography{cats}

\end{document}